\documentclass[10pt,journal,compsoc]{IEEEtran}

\usepackage{algorithm}
\usepackage{algorithmic}
\usepackage{array}
\usepackage{textcomp}
\usepackage{multirow}
\usepackage{stfloats}
\usepackage{url}
\usepackage{booktabs}
\usepackage{verbatim}
\usepackage{graphicx} 
\usepackage{cite}
\usepackage{xcolor}
\usepackage{braket}
\usepackage{amssymb}
\usepackage{amsfonts}
\usepackage{bm}
\usepackage{amsmath}
\usepackage{amsthm}
\usepackage{mathrsfs}
\usepackage{acro}
\usepackage[colorlinks=True,
            linkcolor=blue,
            anchorcolor=green,
            citecolor=blue
            ]{hyperref} 
\usepackage[noabbrev,capitalise,nameinlink]{cleveref}
\usepackage{float}
\usepackage{graphicx}
\usepackage{xcolor}
\usepackage{epstopdf}
\usepackage{multirow}
\usepackage{caption,subcaption}
 \usepackage{enumitem}
\usepackage{etoolbox}
\usepackage{fancyhdr}
\usepackage{ upgreek }
\usepackage{soul}
\usepackage{xcolor}

\makeatletter
\patchcmd{\@algocf@start}
  {-1.5em}
  {0pt}
  {}{}
\makeatother

\usepackage{xpatch}

\usepackage{hyperref}

\makeatletter
\xpatchcmd\SetKwInOut
  {\hangafter=1\parbox[t]}
  {\hangafter=1\justify\parbox[t]}
\makeatother
\usepackage{ragged2e}

\newtheorem{theorem}{Theorem}[section]

\newtheorem{lemma}{Lemma}[section]

\newtheorem{remark}{Remark}[section]
\newtheorem{assumption}{Assumption}[section]

\graphicspath{ {./images/} }

\ifCLASSINFOpdf
\else
\fi

\ifCLASSOPTIONcompsoc
\else
  \usepackage{cite}
\fi

\begin{document}
\title{Tackling Heterogeneity in Quantum Federated Learning: An Integrated Sporadic-Personalized Approach}

\author{
{Ratun Rahman, Shaba Shaon, and Dinh C. Nguyen}
\IEEEcompsocitemizethanks{
\IEEEcompsocthanksitem Ratun Rahman, Shaba Shaon, and Dinh C. Nguyen are with ECE Department,  University of Alabama in Huntsville, Huntsville, AL 35899, USA, emails: (rr0110@uah.edu, ss0670@uah.edu, dinh.nguyen@uah.edu).

}

\thanks{~}}

\IEEEtitleabstractindextext{\justify
\begin{abstract} Quantum federated learning (QFL) emerges as a powerful technique that combines quantum computing with federated learning to efficiently process complex data across distributed quantum devices while ensuring data privacy in quantum networks.  Despite recent research efforts, existing QFL frameworks struggle to achieve optimal model training performance primarily due to inherent heterogeneity in terms of (i) quantum noise where current quantum devices are subject to varying levels of noise due to varying device quality and susceptibility to quantum decoherence, and (ii) heterogeneous data distributions where data across participating quantum devices are naturally non-independent and identically distributed (non-IID). To address these challenges, we propose a novel integrated sporadic-personalized approach called SPQFL that simultaneously handles quantum noise and data heterogeneity in a single QFL framework. It is featured in two key aspects: (i) for quantum noise heterogeneity, we introduce a notion of sporadic learning to tackle quantum noise heterogeneity across quantum devices, and (ii) for quantum data heterogeneity, we implement personalized learning through model regularization to mitigate overfitting during local training on non-IID quantum data distributions, thereby enhancing the convergence of the global model. Moreover, we conduct a rigorous convergence analysis for the proposed SPQFL framework, with both sporadic and personalized learning considerations. Theoretical findings reveal that the upper bound of the SPQFL algorithm is strongly influenced by both the number of quantum devices and the number of quantum noise measurements. Extensive simulation results in real-world datasets also illustrate that the proposed SPQFL approach yields significant improvements in terms of training performance and convergence stability compared to the state-of-the-art methods.
\end{abstract}


}

\maketitle


\IEEEdisplaynontitleabstractindextext

%
\IEEEpeerreviewmaketitle

\ifCLASSOPTIONcompsoc
\IEEEraisesectionheading{\section{Introduction} \label{sec:introduction}}
\else
\section{Introduction}
 \label{sec:introduction}
\fi

%
%
%
%

\IEEEPARstart{Q}{uantum} machine learning (QML) is an emerging field that integrates quantum computing with machine learning (ML) across distributed quantum devices. This synergy enables QML to process complex data at unprecedented speeds, addressing challenges that traditional ML methods struggle with \cite{marshall2023high, haug2023quantum}. In conventional QML frameworks, a central server collects and processes data, raising significant concerns related to data privacy and high communication overhead, particularly for large-scale complex data \cite{chehimi2022quantum}. These limitations make it impractical for devices to transmit their local data to a centralized server continuously. Federated learning (FL) has been introduced to address this problem by enabling devices to train ML models locally and share only trained models with the central server for global aggregation for model privacy preservation. Naturally, FL combined with QML is used to create a new learning paradigm called quantum federated learning (QFL), where quantum devices, known as noisy intermediate-scale quantum (NISQ) devices, collaborate with a central quantum server to train a shared quantum ML model in a quantum network. This approach enhances data security and significantly reduces communication latency through quantum protocols \cite{huang2022quantum, nguyen2025quantum, ren2025toward}. Despite such promising research efforts \cite{bhatia2024communication, javeed2024quantum, qiao2024transitioning}, two practical challenges arise due to inherent heterogeneity, i.e., \textit{heterogeneous quantum noise} and   \textit{non-independent and identically distributed (non-IID) quantum data}.

\textbf{1. Quantum Noise:} NISQ devices are susceptible to quantum noise, resulting from unexpected interactions and hardware defects during quantum operations. Additionally, factors such as hardware decoherence, electromagnetic interference, temperature fluctuations, and imperfect qubit control in quantum networks further exacerbate this noise \cite{zou2024spatially}. These diverse and unpredictable noise sources significantly affect training performance, degrading the stability and reliability of quantum learning algorithms. This inherent unpredictability adds additional complexity to the QFL framework. Effective quantum noise-controlling strategies could enhance the coherence between qubits and quantum layers, boosting computational performance and enabling more reliable, large-scale quantum systems. Quantum noise originates from two main sources, i) quantum gates: where interactions with the environment or defects in the gate operations induce decoherence. and ii) quantum layers: where many gate operations are performed simultaneously, defects such as decoherence and gate defects accumulate and damage qubits collectively. This leads to a critical question: a) \textit{How can inherent quantum noise be controlled to stabilize QML model training in QFL?} 

\textbf{2. Non-IID Quantum Data:} Another significant challenge that QFL faces is its difficulty in handling non-IID quantum data. This data heterogeneity creates high variance among individual devices in the quantum data distribution \cite{11239475, sajadimanesh2025nr}, which often restricts the effectiveness of the globalization process, resulting in suboptimal global model performance. Data heterogeneity and non-IID data distribution lead to concerns about poor convergence and accuracy since devices can perform slightly different learning tasks \cite{pokharel2025quantum}. In real-life scenarios, different devices can also have varying processing power, computational abilities, and resources \cite{11283618, 11314585}. Moreover, quantum data inherently exhibits heterogeneity due to unique quantum encoding and processing characteristics. Effectively managing heterogeneous data in QFL may enable a broader range of devices to participate, enhancing the system's robustness and boosting overall performance. This raises another fundamental question: b) \textit{How can heterogeneous data distribution in non-IID settings be effectively addressed in QFL?} \textbf{This paper aims to answer these fundamental questions through the following contributions.}

\begin{itemize}
    \item We propose \textit{SPQFL}, a novel integrated sporadic-personalized quantum federated learning framework that simultaneously addresses quantum noise and data heterogeneity in NISQ devices within a QFL environment. Our scheme improves local training on quantum devices, ensuring stable global model performance, and is the first to explore sporadic personalized learning in heterogeneous QFL in Section~\ref{Sec: method}.
    \item We introduce sporadic learning to adapt local model updates based on quantum noise levels across NISQ devices in Subsection~\ref{subsec: spo} and apply regularization-based personalization to mitigate overfitting from non-IID data, enhancing global model generalization and convergence stability in heterogeneous QFL environments in Subsection~\ref{subsec: personalized}.
    \item We conduct a rigorous convergence analysis of the proposed SPQFL framework, showing that increasing the number of devices and quantum measurement shots improves training stability and performance in Subsection~\ref{ConvergenceAnalysis} and detailed proof in Section~\ref{Sec: proof}. Moreover, extensive experiments based on MNIST, FashionMNIST, CIFAR-100, and Caltech-101 datasets demonstrate that SPQFL outperforms state-of-the-art benchmarks by up to $6.25\%$ in Section~\ref{Sec: Experiments}. 
\end{itemize}




\begin{figure}
    \centering
    \includegraphics[width=0.99\linewidth]{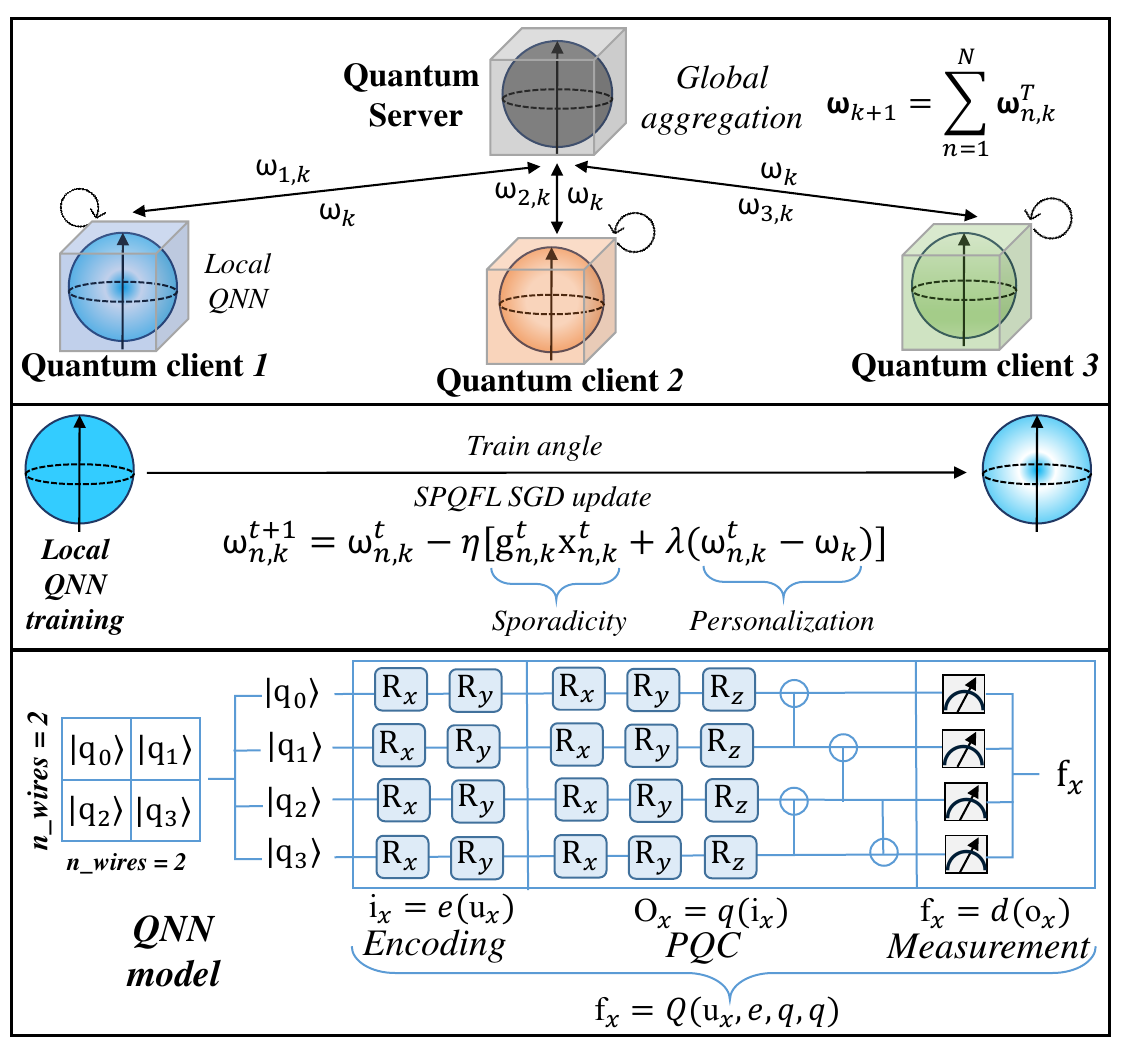}
    \caption{\footnotesize Proposed \textit{SPQFL} architecture where a set of distributed quantum devices collaborate with a quantum server to train a shared QML model. The proposed framework encompasses two key aspects, namely sporadic learning and personalized learning. The sporadicity term captures noise heterogeneity, while the personalization term addresses data heterogeneity.  }
    \label{fig: overview}
\end{figure}

\section{Related Works}
\subsection{Quantum Machine Learning}
QML leverages quantum properties to accelerate classical ML techniques. In comparison to classical algorithm counterparts, QML can achieve exponential or quadratic acceleration in both supervised learning, where models are trained using labeled data to predict outcomes or categorize incoming data, and unsupervised learning, which allows models to uncover patterns and structures in unlabeled data without explicit direction. 
QML's unique qualities make it an effective choice for dealing with complicated and large-scale datasets that challenge standard ML approaches, as authorized and demonstrated in \cite{marshall2023high, haug2023quantum}. Further studies in \cite{oh2020tutorial, sebastianelli2021circuit} proposed a hybrid QML system where devices initially train on QML and then use classical ML techniques such as convolutional neural networks (CNN) for data processing and model training. Although it performs better than QML, hybrid QML requires very complicated ML models, which require significant resources and processing power. 

\subsection{Quantum Federated Learning}
Despite the advantages, QML has two notable drawbacks: 1) it generally relies on data sharing and centralized data collection, which raises substantial privacy concerns, and 2) it demands significant resources to process quantum transformation for large and complex data and poses stability challenges. QFL tackles the inherent limitations of QML by integrating the FL framework into the QML framework. In QFL, NISQ devices locally train their QML models on their respective stored data in a quantum network. In every global round, every device independently updates and generates local model and transmits the local model to a centralized server. The server collects all local models, aggregates them to create an updated global model, and broadcasts the updated global model to all devices for the next global round of training \cite{chehimi2022quantum}. Every device again retains the updated global model, combining it with local data, and updates the model accordingly with a learning rate. Using decentralized data in a collaborative approach and sharing local models, QFL can provide enhanced performance over traditional QML \cite{huang2022quantum} while also improving feasibility and robustness \cite{ren2025toward}. Additionally, QFL is more efficient in communication with large and complex datasets than the QML approach. The reduced communication overhead decentralized approach lowers bandwidth usage and communication frequency, resulting in improved scalability \cite{bhatia2024communication}. Furthermore, by utilizing the properties of quantum computing and quantum cryptography, QFL can provide better security and privacy to the local models than classical FL. 

\subsection{Noise aware Learning}
Recent works in QML have begun to address noise-related concerns.  Liu et al. \cite{liu2024quantum} and Temme et al. \cite{temme2017error} introduced noise-resistant circuit execution methods. Schuld et al. \cite{schuld2021fault} studied fault-tolerant QML and discovered that quantum noise greatly reduces model accuracy, particularly in variational methods.  Sajadimanesh et al. \cite{sajadimanesh2025nr} examined the behavior of noisy QNNs and underlined the importance of in-training mitigation over post-hoc correction.

In FL contexts, Li et al. \cite{li2024privacy} presented privacy-preserving QFL under simple noise channels, but did not include noise in learning dynamics.  There is still a research gap in explicitly simulating quantum noise as part of the optimization process.  Adaptive weighting schemes in classical FL, such as SpoQFL \cite{rahman2025sporadic}, SCAFFOLD \cite{karimireddy2020scaffold}, and Agnostic FL \cite{cortes2020agnostic}, aim to address client unreliability. However, these mechanisms do not directly translate to quantum settings due to fundamentally different noise characteristics and quantum gradient estimation methods (e.g., parameter-shift rule).

\textbf{Despite these research efforts, data and noise heterogeneity issues in QFL have not been addressed, which motivates us to conduct this study.} \textcolor{black}{The proposed framework is illustrated in Fig.~\ref{fig: overview} where there are 3 quantum devices (clients), each training a QNN model using encoding, PQC, and measurement and finally updating local model with i) sporadicity and ii) personalization.}


\noindent

\section{Proposed Integrated Sporadic-Personalized QFL (\textit{SPQFL}) Methodology} \label{Sec: method}

\textbf{QNN Model.} Quantum Neural Networks (QNNs) are variational models that use quantum computing concepts to perform learning tasks by altering quantum states.  There are three main components in a typical QNN: (i) a quantum encoding layer, which uses parameterized rotation gates (e.g., RX, RZ, or ZZ) to embed classical input features into quantum states; (ii) a variational layer, which is made up of trainable quantum circuits with entangling gates (e.g., CNOT or CZ), which forms the core of the model where it primarily learns; and (iii) a measurement layer, which measures quantum observables, usually in the Pauli-Z basis, to extract classical output for tasks such as classification or regression. The training procedure uses a hybrid quantum-classical framework, with quantum devices evaluating the parameterized circuit and classical optimizers updating parameters using gradients or heuristic algorithms.  This hybrid structure makes QNNs ideal for integration into federated frameworks, allowing for lightweight on-device computing while potentially providing collaborative advantages in high-dimensional feature spaces.

\subsection{Local Model Training}
\textbf{Quantum encoding.} A QNN model is trained by first converting classical input into quantum states, also known as quantum encoding. In our approach, to convert classical data into quantum states, each client locally trains a QNN using angular encoding.  Angular encoding takes an input vector \(w \in \mathbb{R}^{d} \) and translates each feature \(w_i \) to a rotation around the Y-axis on the \(i \)-th qubit.  Unitary encoding is defined as
\begin{equation}
    U_{\text{enc}}(w) = \bigotimes_{i=1}^{d} RY(w_i),
\end{equation}
where the rotation gate around the Y-axis for feature \( w_i \) is represented by \(RY(w_i) = \exp(-i w_i Y / 2) \).  It converts the original state \( |0\rangle^{\otimes D} \) into the encoded state as

\begin{equation}
    |\psi_{\text{enc}}(w)\rangle = U_{\text{enc}}(w)|0\rangle^{\otimes D},
\end{equation}
where the number of qubits is equal to the input dimension \(D = d\).

\textbf{PQC training.} The encoded quantum state is subsequently processed by a parameterized quantum circuit (PQC) constructed of $L$ layers of trainable unitary gates.  Each layer consists of single-qubit rotation gates and two-qubit entanglement gates, such as CNOTs.  The entire variational transformation is represented as
\begin{equation} \label{eq:PQC_unitary_transformation}
|\psi_{\text{out}}(w, \omega)\rangle = U(\omega)U_{\text{enc}}(w)|0\rangle^{\otimes D},
\end{equation}
where \( \omega \in \mathbb{R}^{p} \) is the set of trainable parameters.

\textbf{Measurements.} \textcolor{black}{$n \in \mathcal{N}$ denote the client index and $k \in \mathcal{K}$ denote the class index.} The model output is obtained by measuring the expected value of a Hermitian observable $O$, commonly a Pauli-Z operator on a chosen qubit as
\begin{equation}
    f(w, \omega) = \langle\psi_{\text{out}}(w,\omega)|O|\psi_{\text{out}}(w,\omega)\rangle.
\end{equation}

In practice, however, this value is estimated with \(M \) measurement shots as
\begin{equation}
    \hat{f}_{n,k}(\omega) = \frac{1}{M} \sum_{j=1}^{M} H_j,
\end{equation}
where \(H_j \in \{-1, +1\} \) are the measurement results.
The categorical cross-entropy loss is used for training: \begin{equation}
L_{n,k}(\omega) = \ell(y, \hat{f}_{n,k}(\omega)),    
\end{equation} 
where \(y \) true label and $\hat{f}_{n,k}(\omega)$ is the prediction.
The parameter-shift rule is used to estimate gradients in order to optimize the parameters \( \omega \):
\begin{equation}\label{eq:parameter_shift_gradients}
[\hat{g}_{n,k}^{t}]_{d} = \frac{1}{2} \left( \langle\hat{H}\rangle_{|\psi(\omega+\frac{\pi}{2} e_{d})\rangle} - \langle\hat{H}\rangle_{|\psi(\omega-\frac{\pi}{2} e_{d})\rangle} \right),
\end{equation}
where the standard basis vector pointing in the direction of the \( d \)-th parameter is \( e_d \).
Finally, we use gradient descent to update parameters $\omega$ as
\begin{equation}\label{eq:gradient_update}
\omega_{n,k}^{t+1} = \omega_{n,k}^{t} - \eta \hat{g}_{n,k}^{t},
\end{equation}
where \( \eta \) is the learning rate. The parameters \( \omega_{n,k}^{T} \) are sent to the server for federated aggregation after \( T \) local training steps.

\subsection{Effects of Quantum Noise}
Quantum noise has a major effect on QNN training performance.  Quantum devices, particularly Noisy Intermediate-Scale Quantum (NISQ), are extremely sensitive to noise caused by decoherence, gate infidelities, and environmental disturbances.  These noise sources affect quantum states during QNN operation, raising challenges to model convergence, stability, and reproducibility.

In the ideal noise-free scenario, the parameterized quantum circuit (PQC) converts the initial ground state \(|0\rangle \) into a pure quantum state as 
\begin{equation}
    |\Psi(\omega)\rangle = U(\omega) |0\rangle,
\end{equation}
where  the composite unitary \(U (\omega) = \prod_{d=1}^{D}  U_d(w_d) V_d \) is made up of parameterized gates \( U_d(w_d) = \exp \left(-i \frac{w_d}{2} G_d \right) \), in which \( G_d \) is a Pauli string generator (e.g., \( I, X, Y, Z \)), and \( V_d \) is a fixed unitary transformation that is distinct from \( w_d \). 

{\color{black} The \textit{Pauli noise channel} is a popular model for simulating noise. It introduces probabilistic errors on a density matrix $\rho_{ideal}$ (ideal noiseless state of the circuit) to quantum gates as
\begin{equation}
\mathcal{E}(\rho_{\text{ideal}})
= (1 - \epsilon) \rho_{\text{ideal}}
+ \epsilon \sum_{j} E_j \rho_{\text{ideal}} E_j^{\dagger},
\end{equation}
where \(\epsilon \) is the gate error probability and \(E_j \) are Pauli operators (e.g., \(X, Y, Z \)) used stochastically. The ensuing mixed state $\mathcal{E}(\rho_{\text{ideal}})$ reflects the quantum system's noisy development.}

{\color{black}
Quantum noise has a comparable effect on the empirical estimate of the model's loss function. 
In \eqref{eq:parameter_shift_gradients}, each model output $\hat{f}_{n,k}(\omega)$ is produced from a finite number of noisy measurements, therefore the corresponding loss is stochastic. 
The noisy empirical loss can therefore be expressed as
\begin{equation}
    \hat{L}_{n,k}(\omega) 
    = \ell\!\big(y,\, \hat{f}_{n,k}^{\text{noisy}}(\omega)\big)
    = \ell\!\left(y,\, \frac{1}{M}\sum_{j=1}^{M} H_{j,k}^{\text{noisy}}\right),
\end{equation}
where $H_{j,k}^{\text{noisy}}\!\in\!\{-1,+1\}$ represents shot outcomes influenced by gate and measurement noise. 
Residual noise causes changes in $\hat{L}_{n,k}(\omega)$ even after averaging over $M$ shots, causing variance in gradient estimations.
}
Under noise, the parameter-shift gradient in \eqref{eq:parameter_shift_gradients} is expressed as
\begin{equation}\label{eq:gradient_with_noise}
\hat{g}_{n,k}^t = \nabla f(\omega_{n,k}) + \xi_{n,k}^t,
\end{equation}
where \( \nabla f(\omega_{n,k}) \) is the true (noise-free) gradient, and \( \xi_{n,k}^t \) is a noise-integrated perturbation. The variance of this noise term is bounded by
\begin{equation}\label{eq:variance_of_noise}
\text{Var}(\xi_{n,k}^t) \leq \frac{\nu N_h D \, \text{Tr}(H^2)}{2M},
\end{equation}
where \textcolor{black}{ $\nu$ is a hardware-dependent constant based on the overall noise rate $\epsilon$.
 $\epsilon$ is the average Pauli error probability per gate (or per layer) and can be calculated experimentally from device calibration data, such as from IBM Quantum backends or Rigetti systems, where single- and two-qubit gate error rates are released frequently. 
 The proposed sporadic technique involves updating $\epsilon$ at the start of each communication cycle with the most recent calibration to adaptively weight noisy gradients.}

The convergence is impacted over time by this accumulation of noise.  Following \( T \) communication rounds, the noise-induced convergence gap's upper bound as
\begin{equation}\label{eq:error_bound_of_noise_in_loss}
\xi_t = \mathbb{E}[L(\omega_T)] - L^* \leq (1 - \eta \mu)^T \left(\mathbb{E}[L(\omega_0)] - L^*\right) + \frac{1}{2} \frac{\eta L V}{\mu},
\end{equation}
where \( L^* \) is the optimal loss, \( \eta \) is the learning rate, \( \mu \) is the strong convexity parameter, \( L \) is the Lipschitz constant, and \( V \) denotes the variance of noise. 
\textcolor{black}{The bound in \eqref{eq:error_bound_of_noise_in_loss} reflects standard noisy-gradient evaluations for stochastic optimization, and its quantum extension parallels previous results on convergence under noisy gradient estimation in variational quantum algorithms.}

\subsection{Proposed Sporadic Learning Method} \label{subsec: spo}
To address the instability caused by quantum noise in NISQ-era devices, we present a \textit{sporadic learning} mechanism that adjusts the contribution of each local update based on its noise level. The main idea is straightforward: instead of allowing each local device to participate equally in every training step, we force each device to scale down its update if it is currently encountering a lot of noise.  In this way, noisy updates have a lower influence on the global model, while accurate updates can contribute constantly. This technique allows each client \(n \) to calculate an efficient local model \(\boldsymbol{w}_n \), which accounts for quantum disturbances while reducing communication cost.

During each communication round \( k \), and for each local iteration \( t \), the local update rule for device \( n \) is defined as
\begin{equation} \label{eq:local_sgd_sporadic}
\boldsymbol{w}_{n,k}^{t+1} = \boldsymbol{w}_{n,k}^t - \eta \, g_{n,k}^t \, x_{n,k}^t,
\end{equation}
where \( \eta \) is the learning rate, \( g_{n,k}^t \) is the local stochastic gradient, and \( x_{n,k}^t \in [0,1] \) is a noise-aware weight that controls the magnitude of the update. The gradient term incorporates the influence of quantum noise as:
\begin{equation} \label{eq:gradient_with_noise_term}
g_{n,k}^t = \nabla f_n(\boldsymbol{w}_{n,k}^t) + \xi_{n,k}^t,
\end{equation}
where \( \nabla f_n(\cdot) \) is the true gradient of the local loss function and \( \xi_{n,k}^t \) represents the additive noise induced by quantum operations, as previously defined in \eqref{eq:gradient_with_noise}.

The sporadic learning strategy's novelty is the modulation term \(x_{n,k}^t \), which probabilistically reduces the update intensity based on the observed noise level.  The computation is described as
\begin{equation} \label{eq:noise_control_factor}
x_{n,k}^t = \exp(-\gamma \, \xi_{n,k}^t),
\end{equation}
where the adjustable hyperparameter \(\gamma > 0 \) affects the sensitivity of the decay function.  In noisy situations, higher values of \(\gamma \) result in a sharper decrease in update strength.

Intuitively, when the noise is minimal, \(x_{n,k}^t \approx 1 \), it allows a full local update.  In contrast, when noise is large, \(x_{n,k}^t \ll 1 \), suppresses faulty updates.  This approach assures that local models on very noisy devices do not have an adverse impact on global aggregation, while still allowing participation without adding additional expense to the server or quantum client.  Importantly, the proposed approach is lightweight and completely decentralized, depending just on local noise measurements and update scaling.  As a consequence, it improves convergence stability in the presence of noise without the need for hardware-level error correction or repetitive circuit execution.

Therefore, the sporadic learning approach is simple yet powerful.  Instead of excluding noisy clients or forcing them to address their noise problems, which may be impractical, we \textit{mitigate their influence}.  If a quantum device generates a noisy update, we minimize its contribution proportionately, without requiring additional communication or server-side adjustments.  Eventually, even in a federated network of noisy quantum clients, this makes the global model more accurate and stable over time.

 \subsection{Personalized Learning For Handling Quantum Data Heterogeneity} \label{subsec: personalized}
Different quantum devices or devices frequently encounter unique noise patterns, calibration errors, and hardware restrictions, resulting in quantum data heterogeneity. This heterogeneity makes it difficult to apply a one-size-fits-all global model efficiently. Personalized learning with model regularization addresses this issue by allowing each device to fine-tune its local models to better fit its unique data and noise characteristics. In our approach, regularization ensures that devices that perform suboptimal performance during local training continue training for further epochs locally rather than instantly sending their model updates to the global server. This adaptive technique mitigates the effects of quantum noise and heterogeneity, ensuring that the global model receives higher-quality updates, thereby boosting both global convergence and local performance.
 
In QFL, the goal is to minimize the loss value of all the devices, which is expressed as
\begin{equation} \label{eq: basicFL}
    \min_{\boldsymbol{w} \in {\rm I\!R}^d} F(\boldsymbol{w}) := \frac{1}{N} \sum_{n=1}^{N} f_n(\boldsymbol{w}),
\end{equation}
where $f_n(\boldsymbol{w})$ is the local loss function computed by the $n^{th}$ device using its own data. In any conventional QFL framework, the local model is then updated using sporadic SGD as \eqref{eq:local_sgd_sporadic}, each device eventually shares its updated model with the server for equal contribution to the global model. In contrast to this traditional approach, \textcolor{black}{in our SPQFL approach}, we first calculate the local loss value $f_n (\mathbf{\boldsymbol{w}}_{n,k})$ and compare it with the preceding epoch's global loss $F(\boldsymbol{w}_{k-1}^{\text{global}})$. Specifically, we calculate 
\begin{equation}
    f_n (\mathbf{\boldsymbol{w}}_{n,k}) = \langle \psi_{\text{out}}(\mathbf{\boldsymbol{w}_{n,k}}, n) |\, O \,| \psi_{\text{out}}(\mathbf{\boldsymbol{w}_{n,k}}, n) \rangle.
\end{equation}
If the following inequality holds true
\begin{equation*} \label{eq: conditional_epochs}
    f_n(\boldsymbol{w}_{n,k}) > F(\boldsymbol{w}_{k-1}^{\text{global}}),
\end{equation*}
the device performs additional local training epochs. The number of extra epochs can be dynamically set based on the difference between $f_n(\boldsymbol{w})$ and $F(\boldsymbol{w}_{k-1}^{\text{global}})$, or capped at a maximum threshold. 

During each local epoch, the device updates its model parameters based on personalized learning as
\begin{equation} \label{eq: local_sgd_FL_quantum}
    \boldsymbol{w}_{n,k}^{t+1} = \boldsymbol{w}_{n,k}^{t} - \eta \left[g_{n,k}^t + \lambda(\boldsymbol{w}_{n,k}^{t} - \boldsymbol{w}_{k})\right],
\end{equation}

\noindent
By combining sporadic learning and personalized learning in equation \eqref{eq:local_sgd_sporadic} and \eqref{eq: local_sgd_FL_quantum}, we come up with the final model update rule as
\begin{equation}
    \boldsymbol{w}_{n,k}^{t+1} = \boldsymbol{w}_{n,k}^{t} - \eta \left[g_{n,k}^t x_{n,k}^t + \lambda(\boldsymbol{w}_{n,k}^{t} - \boldsymbol{w}_{k})\right], \label{eqnlabelnew1}
\end{equation} \noindent 

\noindent
where $\lambda >0$ regulates the interpolation of global and individual models. Finally, each device communicates its revised model parameters $\boldsymbol{w}_{n,k} = \boldsymbol{w}_{n,k}^T$ to the central server.

\textbf{Local Model Aggregation.} The central server collects model updates from all devices to aggregate and create the new global model:
\begin{equation} \label{eq: fedavg_personalized}
    \boldsymbol{w}_{k+1} = \frac{1}{N} \sum_{n=1}^{N} \boldsymbol{w}_{n,k}^{T}.
\end{equation}

\noindent
This global aggregation takes into account any individualized updates from quantum devices that complete further epochs, resulting in a more adaptive global model. The proposed personalized method enables each device to better optimize its specific data distribution while still contributing to the global model. PQFL intends to increase overall convergence and individual device performance by dynamically altering the number of local epochs based on the global model's performance.

\begin{algorithm}[ht]
\small
\caption{Sporadic Quantum Federated Learning}
\label{alg:sporadic_fl}
\begin{algorithmic}[1]
\STATE \textbf{Input:} Learning rate \( \eta \), number of rounds \( K \), local steps \( T \), noise sensitivity \( \gamma \), number of clients \( N \)
\STATE \textbf{Initialize:} Global model \( \omega_0 \)
\FOR{each round \( k = 1 \) to \( K \)} \label{line:rounds}
    \STATE Server sends \( \omega_k \) to selected clients
    \FOR{each client \( n \in \mathcal{S}_k \) \textbf{in parallel}} \label{line:client_loop}
        \STATE Initialize \( \omega_{n,k}^0 \gets \omega_k \)
        \FOR{local step \( t = 0 \) to \( T-1 \)} \label{line:local_loop}
            \STATE Compute gradient with noise: \( g_{n,k}^t = \nabla f_n(\omega_{n,k}^t) + \xi_{n,k}^t \) \label{line:gradient}
            \STATE Compute noise-aware weight: \( x_{n,k}^t = \exp(-\gamma \, \xi_{n,k}^t) \) \label{line:weight}
            \STATE Update local model using spodaric and personalized learning in \eqref{eqnlabelnew1} \label{line:update}
        \ENDFOR
        \STATE Client sends \( \omega_{n,k}^T \) to server
    \ENDFOR
    \STATE Server aggregates models using model regularization \eqref{eq: fedavg_personalized}
\ENDFOR
\STATE \textbf{Return:} Optimal global model \( \omega_K \)
\end{algorithmic}
\end{algorithm}

We summarize our proposed \textit{SPQFL} methodology explained above in Algorithm \ref{alg:sporadic_fl}. The server starts by initializing a global model and goes through \(K \) communication rounds (Line \ref{line:rounds}).  In each round, a subset of customers is chosen and given the current global model.  Each client does local training in parallel (Line \ref{line:client_loop}).  For each \(T \) local step (Line \ref{line:local_loop}), the client first computes its local gradient, which incorporates noise from quantum computing (Line \ref{line:gradient}). To limit the influence of noise, the client computes a noise-aware scaling factor using an exponential decay function (Line \ref{line:weight}).  This component influences the gradient step, resulting in a noise-controlled model update with model personalization (Line \ref{line:update}).  Following local training, each client sends its final model to the server, which combines the results to update the global model.  This sporadic approach enables the system to dynamically suppress unreliable updates, increasing resilience to noise.

\color{black}

\section{Convergence Analysis} \label{ConvergenceAnalysis}

To facilitate our convergence analysis, we define the following: 
$\bar{\boldsymbol{w}}_k^t = \frac{1}{N} \sum_{n\in\mathcal{N}} \boldsymbol{w}_{n,k}^t$,
$g_k^t = \frac{1}{N} \sum_{n\in\mathcal{N}} g_{n,k}^{t} = \frac{1}{N} \sum_{n\in\mathcal{N}} \nabla F(\boldsymbol{w}_{n,k}^t,\chi_{n,k}^t)$, $\bar{g}_k^t = \frac{1}{N} \sum_{n\in\mathcal{N}} \nabla F(\boldsymbol{w}_{n,k}^t)$, $\bar{\boldsymbol{x}}_{k}^{t} = \frac{1}{N} \sum_{n \in \mathcal{N}} \boldsymbol{x}_{n,k}^{t}$,
\noindent
where $F$ is a local loss function, and $\chi$ is a sample uniformly chosen from the local dataset. It is apparent that $\bar{\boldsymbol{w}}_{k}^{t+1} = \bar{\boldsymbol{w}}_{k}^{t} - \eta \left[g_{k}^{t}\bar{\boldsymbol{x}}_{k}^t + \lambda \bar{\boldsymbol{w}}_{k}^{t} - \boldsymbol{w}_{k}) \right]$.
\noindent
We also mention that \(
    \mathbb{E}g_{k}^{t} = \bar{g}_{k}^{t},\)
where $\mathbb{E}$ represents function's expectation. In this paper, we focus on the federated quantum convex neural network (QNN) setting which was investigated in \cite{bengio2005convex}, where the proposed \textit{SPQFL} model relies on classical loss functions to optimize and train its gradient. It motivated us to conduct a convergence analysis under a convex loss function setting. To support our analysis, we use the following standard assumptions:
\begin{assumption} \label{Assumption1}
 Each local loss function $F_n$ ($n\in \mathcal{N}$) is $L$-smooth, i.e. $F_n(\boldsymbol{w}') - F_n(\boldsymbol{w}) \leq \langle \boldsymbol{w}'- \boldsymbol{w}, \nabla F(\boldsymbol{w} \rangle + \frac{L}{2} ||\boldsymbol{w}'- \boldsymbol{w}||, \forall \boldsymbol{w}', \boldsymbol{w}$.
 \end{assumption}
 \begin{assumption} \label{Assumption2}
 Each local loss function $F_n$ ($n\in \mathcal{N}$) is $\mu$-strongly convex, i.e. $F_n(\boldsymbol{w}') - F_n(\boldsymbol{w}) \ge \langle \boldsymbol{w}'- \boldsymbol{w}, \nabla F(\boldsymbol{w} \rangle + \frac{\mu}{2} ||\boldsymbol{w}'- \boldsymbol{w}||, \forall \boldsymbol{w}', \boldsymbol{w}$.
 \end{assumption}
 \begin{assumption} \label{Assump:Variance-gradient}
 The variance of stochastic gradients on local model training at each device is bounded: $\mathbb{E}||\nabla F(\boldsymbol{w}_{n,k}^t,\chi_{n,k}^t) - \nabla F(\boldsymbol{w}_{n,k}^t)||^2 \leq \sigma_g^2$. \label{Assumption3New}
 \end{assumption}
\noindent
The convergence of  \textit{SPQFL} is characterized by the following lemmas:
\begin{lemma} \label{Lemma1}
Let Assumption \ref{Assump:Variance-gradient} hold, the expected upper bound of the variance of the stochastic gradient on local model training is given as $\mathbb{E} ||g_k^t- \bar{g}_k^t||^2 \leq \frac{\sigma_g^2}{N^2}$,
where $\sigma_g^2 = \frac{\nu N_{h} D Tr(H^{2})}{2M}$.
\end{lemma}
\begin{proof}
See Section \ref{ProofLemma1}. \renewcommand{\qedsymbol}{}
\end{proof}
\begin{lemma} \label{Lemma2}
The expected upper bound of the divergence of $\boldsymbol{w}_{n,k}^t$ is given as $\left[ \frac{1}{N}\sum_{n\in\mathcal{N}} \mathbb{E} \Big\Vert\bar{\boldsymbol{w}}_k^t-\boldsymbol{w}_{n,k}^t\Big\Vert^2  \right] \leq \frac{1}{2}\eta T G^2$, for some positive $G$.
\end{lemma}

\begin{proof}
    See Section \ref{ProofLemma2}. \renewcommand{\qedsymbol}{}
\end{proof}

\begin{lemma} \label{Lemma3}
The expected upper bound of $\mathbb{E} \left[||\bar{\boldsymbol{w}}_k^{t+1} - \boldsymbol{w}^*||^2 \right]$ is given as
\begin{equation} 
\begin{aligned}
&\mathbb{E}||\bar{\boldsymbol{w}}_k^{t+1} - \boldsymbol{w}^*||^2 
 \\&\leq \left(1+\eta + \frac{\lambda \eta}{2} + \frac{\lambda}{\eta} + 2\lambda^2 + \bar{\boldsymbol{x}}_k^t \lambda L_p \right)\mathbb{E}||\bar{\boldsymbol{w}}_k^t - \boldsymbol{w}^*||^2 
 \\&+ (1+\frac{1}{2\eta})\mathbb{E}||g_k^t \bar{\boldsymbol{x}}_k^t-\bar{g}_k^t \bar{\boldsymbol{x}}_k^t||^2 
 + \left( 2\lambda + 2\lambda^2 \eta^2 + \lambda\eta \right) \\&\mathbb{E}||\boldsymbol{w}^* - \boldsymbol{w}_k||^2 
 + \bigg(2\lambda \eta^2 (1+\eta) + \bigg(2\lambda \eta^2 \bar{\boldsymbol{x}}_k^t 
 \\&+ 2L_p\bigg(1 + 2\lambda + \lambda \eta + \frac{2\lambda}{\eta} + \frac{1}{2\eta}\bigg)\bar{\boldsymbol{x}}_k^t\bigg) \bigg)
 \\& \times \mathbb{E}\bigg(\frac{1}{N} \sum_{n\in\mathcal{N}}||\bar{\boldsymbol{w}}_k^t -\boldsymbol{w}_{n,k}^t||^2\bigg)
\\&+ \left(2\lambda \eta^2 \bar{\boldsymbol{x}}_k^t + 2L_p\left(1 + 2\lambda + \lambda \eta + \frac{2\lambda}{\eta} + \frac{1}{2\eta}\right)\bar{\boldsymbol{x}}_k^t\right) 
\\&\times \left(\frac{1}{2L_p}+\frac{1}{2\vartheta}\right) ||\nabla F(\bar{\boldsymbol{w}}_{k}^t)||^2. 
\end{aligned}
\end{equation}
\end{lemma}

\begin{proof}
    See Section \ref{ProofLemma3}. \renewcommand{\qedsymbol}{}
\end{proof}

Based on the above lemmas, the convergence bound of the  \textit{SPQFL} is stated in the following theorem.
\begin{theorem} \label{Theorem1}
Let Assumptions 1-3 hold,  then the upper bound of the \textit{SPQFL}'s convergence rate after $K$ global rounds satisfies: 
\begin{equation} \label{equa:final_convergenceIIDnew}
\begin{aligned}
&\mathbb{E}\left[F(\boldsymbol{w}_K)\right] -F^* 
\\&\leq \frac{L}{2(K+L/\mu)}\left[\frac{9\Phi_K}{8\mu^2} + \left( \frac{L}{\mu}+1\right) \mathbb{E}||\boldsymbol{w}_0 - \boldsymbol{w}^*||^2 \right],
\end{aligned}
\end{equation}
where
\begin{align}
    &\Phi_K = \frac{1}{2}\eta \bigg(2\lambda \eta^2 (1+\eta) + \bigg(2\lambda \eta^2 \bar{\boldsymbol{x}}_K \nonumber
    \\&+ 2L_p\bigg(1 + 2\lambda + \lambda \eta + \frac{2\lambda}{\eta} + \frac{1}{2\eta}\bigg)\bar{\boldsymbol{x}}_K\bigg) \bigg)T G^2 \nonumber
\\&\hspace{4em}+ (1+\frac{1}{2\eta})|\bar{\boldsymbol{x}}_K|^2 \sum_{n \in \mathcal{N}}\frac{\nu N_{h} D Tr(H^{2})}{2MN^2}.  \nonumber
\end{align}
\end{theorem}

\begin{proof}
    See Section \ref{ProofTheorem1}. \renewcommand{\qedsymbol}{}
\end{proof}

\begin{remark}
The convergence upper bound achieved in Theorem \ref{Theorem1} shows that the convergence of the \textit{SPQFL} algorithm is strongly influenced by the number of total devices, as well as the number of measurements performed on the PQC.
\end{remark}

\begin{remark}
From \eqref{equa:final_convergenceIIDnew}, it is clear that increasing the number of total devices $N$ can improve the convergence performance of \textit{SPQFL}. Furthermore, using a greater number of quantum measurement shots $M$ in the PQC also contributes to quicker and more efficient convergence within this framework. However, the last component in the equation of $\Phi_K$ represents the sum of the gradient estimate variances from all the participating devices, which captures the noise variance due to quantum shot noise. As a result, increasing device participation in QFL leads to an escalation in collective noise with each global update. Therefore, strategically increasing both the number of clients $N$ and quantum measurement shots $M$ can more effectively enhance the convergence performance. To further elaborate on these theoretical findings, numerical simulations will be performed. As the convergence performance with respect to the number of clients closely aligns with that of classical FL such as in \cite{wei2020federated}, this aspect can be omitted. We will thus focus on simulations on new aspects, i.e., verifying the impacts of the number of measurement shots on \textit{SPQFL} convergence, as shown in Fig.~\ref{fig1} in Section~\ref{subsection-noise}.
\end{remark}

\noindent \textbf{Computational Complexity:}
We now focus on the complexity analysis of the quantum training process. \textcolor{black}{Assuming that Assumption \ref{Assumption1} is valid, i.e., each local loss function $f_n (\cdot)$ is $L$-smooth,} the following inequality can be established:
\begin{align}
    f_{n}(\boldsymbol{w}) \leq f_{n}(\boldsymbol{w'}) + \nabla {f_{n}(\boldsymbol{w}')}^{T}(\boldsymbol{w}-\boldsymbol{w'}) + \frac{L}{2}||\boldsymbol{w}-\boldsymbol{w'}||^{2},
\end{align}
for all $\boldsymbol{w}, \boldsymbol{w'} \in \Theta$. \textcolor{black}{Furthermore, Assumption \ref{Assumption2} holds true, i.e., each local loss function $f_n(\cdot)$ is $\mu$-strongly convex,} there exists a constant $\mu > 0$ and $\mu \leq L$ such that the inequality
\begin{align}
    ||\nabla f_{n}(\boldsymbol{w})||^{2} \geq 2 \mu (f_{n}(\boldsymbol{w}) - {f_{n}}^{*}),
\end{align}
holds for all $\boldsymbol{w} \in \Theta$. In practical applications, the constant $\mu$ is influenced by the number of qubits $n$ and the gate parameters $D$, where typically, a larger quantum circuit is associated with smaller gradient norms. This relationship highlights the impact of quantum circuit complexity on gradient behavior, suggesting that circuits with more qubits and parameters tend to display more subtle and reduced gradient magnitudes. The subsequent theorem establishes a limit on the optimality of the local quantum training output $\boldsymbol{w}^{T}$ after $T$ iterations. The proof of this theorem is inspired by the classical convergence analysis associated with SGD and is elaborated in (\cite{ajalloeian2020convergence}, Theorem 6). 

Assuming $L$-smoothness and $\mu$-strong convexity, for any specified initial point $\boldsymbol{w}^{0}$, the subsequent bound is applicable for any constant learning rate $\eta_{k} = \mu \leq \frac{1}{L}$:
\begin{align}
    \mathbb{E}[f_{n}(\boldsymbol{w}^{T})] - {L}^{*} \leq (1 - \eta \mu)^{T} ([f_{n}(\boldsymbol{w}^{0})] - {L}^{*}) + \frac{1}{2} [\frac{\eta L V}{\mu}],
\end{align}
where $V = \frac{\nu N_{z} D Tr(Z^{2})}{2H}$ is the upper bound of the variance of the gradient estimate, and the expectation is taken over the distribution of the measurement outputs. Moreover, when specifying a desired error threshold $\delta > 0$, for learning rate $\eta = \eta^{\text{shot-noise}} \leq \min\{\frac{1}{L},\frac{\delta \mu}{L V}\}$, a number of iteration, given as
\begin{align}
    T^{\text{shot-noise}} = \mathcal{O} \bigg( \log\frac{1}{\delta} + \frac{V}{\delta \mu} \bigg) \frac{L}{\mu}, \label{eqn26n}
\end{align}
is sufficient to ensure an error $\mathbb{E}[f_{n}(\boldsymbol{w}^{T})-{f_{n}}^{*}] = \mathcal{O}(\delta)$.

Equation \eqref{eqn26n} demonstrates that the local quantum training procedure, utilizing the unbiased gradient estimator, reaches a convergence limit characterized by an error floor approximately $\mathcal{O}(\eta L V)$. By selecting a sufficiently small learning rate $\eta$, this error can be minimized to an arbitrarily low level, ensuring convergence within no more than $T^{\text{shot-noise}}$ iterations. This careful adjustment of the learning rate not only minimizes computational costs but also stabilizes the training process, preventing the need for an excessive number of iterations that could overburden system resources.

\section{Detailed Proofs} \label{Sec: proof}
To substantiate Theorem \ref{Theorem1}, it is necessary to prove Lemmas \ref{Lemma1}, \ref{Lemma2}, and \ref{Lemma3}. Hence, in this section, we provide proofs for the lemmas and theorems outlined in the preceding section.

\subsection{Proof of Lemma \ref{Lemma1}} \label{ProofLemma1}

\noindent
From Assumption \ref{Assump:Variance-gradient}, we have
\begin{align}
&\mathbb{E} ||g_k^t- \bar{g}_k^t||^2 = \mathbb{E} \Big\Vert \frac{1}{N} \sum_{n\in\mathcal{N}} \left(\nabla F(\boldsymbol{w}_{n,k}^t,\chi_{n,k}^t) - \nabla F(\boldsymbol{w}_{n,k}^t) \right)\Big\Vert^2 \nonumber \\ 
&=\frac{1}{N^2}   \sum_{n\in\mathcal{N}} \mathbb{E} \Big\Vert\left(\nabla F(\boldsymbol{w}_{n,k}^t,\chi_{n,k}^t) - \nabla F(\boldsymbol{w}_{n,k}^t) \right)\Big\Vert^2 \leq \frac{\sigma_g^2}{N^2}. \label{eqn15new}
\end{align}
In \eqref{eqn15new}, the term $\nabla F(\boldsymbol{w}_{n,k}^t,\chi_{n,k}^t)$ represents the estimated stochastic gradient, while $\nabla F(\boldsymbol{w}_{n,k}^t)$ refers to the true gradient in the context of our SPQFL framework. We can decompose the gradient estimator $\nabla F(\boldsymbol{w}_{n,k}^t,\chi_{n,k}^t)$ as $\nabla F(\boldsymbol{w}_{n,k}^t,\chi_{n,k}^t) = \nabla F(\boldsymbol{w}_{n,k}^t) + \xi_{n,k}^{t}$, with shot noise $\xi_{n,k}^{t}$ satisfying the conditions $\mathbb{E} \xi_{n,k}^{t} = 0$ and $\text{var}(\xi_{n,k}^{t}) = \mathbb{E}||\nabla F(\boldsymbol{w}_{n,k}^t,\chi_{n,k}^t) - \nabla F(\boldsymbol{w}_{n,k}^t)||^{2}$. Hence, $\sigma_g^2$ \eqref{eqn15new} represents the noise variance, $\text{var}(\xi_{n,k}^{t})$, accumulated on the global model from all the clients at global round $k$.  

We now determine the specific SPQFL parameters that shape the noise variance. We fix the indices related to global and local iteration $k$ and $t$, consequently dropping them from the notations temporarily. Let $X_{d,\pm} = {\langle \hat{H} \rangle}_{\lvert \Psi(\theta \pm \frac{\pi}{2} e_{d}) \rangle} - {\langle H \rangle}_{\lvert \Psi(\theta \pm \frac{\pi}{2} e_{d}) \rangle}$ denote the difference between the estimated and true expectation of the observable $H$ under the quantum state ${\lvert \Psi(\theta \pm \frac{\pi}{2} e_{d}) \rangle}$ whose $d^{\text{th}}$ parameter is phase shifted by $\pm \frac{\pi}{2}$. In the following analysis, we use the notation $\lvert \Psi_{d,\pm} \rangle = {\lvert \Psi(\theta \pm \frac{\pi}{2} e_{d}) \rangle}$ for brevity. The variance of the gradient estimate in \eqref{eqn15new} is written as
\begin{align}
    &\text{var}(\xi) = \mathbb{E} \bigg[ \sum_{d=1}^{D} \bigg( \frac{1}{2} ( {\langle \hat{H} \rangle}_{\lvert \Psi_{d,+} \rangle} - {\langle \hat{H} \rangle}_{\lvert \Psi_{d,-} \rangle}) - \frac{1}{2} ( {\langle H \rangle}_{\lvert \Psi_{d,+} \rangle} \nonumber \\
    &- {\langle H \rangle}_{\lvert \Psi_{d,-} \rangle}) \bigg)^{2} \bigg] 
    = \sum_{d=1}^{D} \frac{1}{4}  \bigg( \mathbb{E}[X_{d,+}^{2}] - \mathbb{E}[X_{d,-}^{2}] \bigg), \label{eqn61}
\end{align}
where the expectation is taken with respect to the $M$ measurements of the quantum states ${\lvert \Psi(\theta + \frac{\pi}{2} e_{d}) \rangle}$ and ${\lvert \Psi(\theta - \frac{\pi}{2} e_{d}) \rangle}$ for $d = 1, 2, \dots, D$. Hence, the random variables $X_{d,+}$ and $X_{d,-}$ are independent for $d = 1, 2, \dots, D$, which results in the equality in \eqref{eqn61}. It is to note that the expectation $\mathbb{E}[X_{d,+}^{2}]$ is equal to the variance $\text{var}({\langle \hat{H} \rangle}_{\lvert \Psi_{d,+} \rangle})$ of the random variable ${\langle \hat{H} \rangle}_{\lvert \Psi_{d,+} \rangle}$. Let $Y$ be the random variable that defines the index of the measurement of the observable $H$. Therefore, $H = h_{Y}$ represents the corresponding measurement output. We denote the Bernoulli random variable as $W_{y} = \mathbb{I}\{Y = y\}$ determining whether $Y = y (W_{y} = 1)$ or not $(W_{y} = 0)$. We also mention that the quantum measurements are i.i.d., and thus it follows from the definition of expectation of ${\langle \hat{H} \rangle}_{\lvert \Psi_{d,+} \rangle}$ that $\mathbb{E}[X_{d,+}^{2}] = \frac{1}{M} \text{var}\bigg( \sum_{y=1}^{N_{h}} h_{y} W_{y} \bigg)
    \leq \frac{N_{h}}{M} \bigg(\sum_{y=1}^{N_{h}} h_{y}^{2}\bigg) v = \frac{N_{h} \operatorname{Tr}(H^{2})}{M} v$,
The variance of the Bernoulli random variable $W_{y}$ is computed as $\text{var}\bigg(W_{y}\bigg) = \mathbb{E}\bigg[W_{y}^{2}\bigg] - \bigg(\mathbb{E}\bigg[W_{y}\bigg]\bigg)^{2} = v\bigg(p(y|\theta + e_{d} \frac{\pi}{2})\bigg)$, where $v(x) = x (1-x)$ for $x \in (0,1)$. The last yields from the definition of the quantity $v$. In a similar way, it can be shown that the following inequality holds $\mathbb{E}[X_{d,-}^{2}] \leq \frac{N_{h} \operatorname{Tr}(H^{2})}{M} v$. We can write while bringing the omitted indices back $\text{var}(\xi_{n,k}^{t}) \leq \frac{\nu N_{h} D Tr(H^{2})}{2M}$. Taking sum of $N$ clients, we have $\text{var}(\xi_{k}^{t}) = \sigma_g^2 \leq \sum_{n \in \mathcal{N}} \frac{\nu N_{h} D Tr(H^{2})}{2M}$, concluding the proof.

\subsection{Proof of Lemma \ref{Lemma2}} \label{ProofLemma2}

\noindent
We know that in every global communication round, each client performs $T$ rounds of local SGDs where there always exists $t' \leq t$ such that $t-t'\leq T$ and $\frac{1}{N} \sum_{n \in \mathcal{N}} \boldsymbol{w}_{n,k}^{t'} = \bar{\boldsymbol{w}}_k^{t'}$, $\forall n \in \mathcal{N}$. By using the fact that $\mathbb{E}||X-\mathbb{E}X||^2 = ||X||^2 - ||\mathbb{E}X||^2$ and $\bar{\boldsymbol{w}}_k^t =\mathbb{E}\boldsymbol{w}_{n,k}^t$, we have $\frac{1}{N}\sum_{n\in\mathcal{N}} \mathbb{E} \Big\Vert\bar{\boldsymbol{w}}_k^t-\boldsymbol{w}_{n,k}^t\Big\Vert^2  
=  \frac{1}{N}\sum_{n\in\mathcal{N}}\mathbb{E}\Big\Vert\left(\sum_{t=t'}^{t-1} \eta \nabla F(\boldsymbol{w}_{n,k}^t,\chi_{n,k}^t) \right) \Big\Vert^2 \leq \frac{1}{N}\sum_{n\in\mathcal{N}}\mathbb{E} \Big\Vert\left(\sum_{t=1}^{t-t'} \eta \nabla F(\boldsymbol{w}_{n,k}^t,\chi_{n,k}^t) \right) \Big\Vert^2$,
where the last inequality holds since the learning rate $\eta$ is decreasing. Assuming that $\eta^{t'} \leq \frac{1}{\sqrt{2}}\eta$ and $||\nabla F(\boldsymbol{w}_{n,k}^t,\chi_{n,k}^t)||^2 \leq G^2$ for positive constant $G$, we have 
\begin{equation} 
\begin{aligned}
\frac{1}{N}\sum_{n\in\mathcal{N}}\mathbb{E} \Big\Vert\bar{\boldsymbol{w}}_k^t-\boldsymbol{w}_{n,k}^t\Big\Vert^2 \leq \frac{1}{N}\sum_{n\in\mathcal{N}} {(\eta^{t'})}^2T G^2 \leq \frac{1}{2}\eta T G^2.
\end{aligned}
\end{equation}

\subsection{Proof of Lemma \ref{Lemma3}} \label{ProofLemma3}

\noindent
From the SGD update rule of the training $\bar{\boldsymbol{w}}_{k}^{t+1} = \bar{\boldsymbol{w}}_{k}^{t} - \eta \left[g_{k}^{t}\bar{\boldsymbol{x}}_{k}^t + \lambda (\bar{\boldsymbol{w}}_{k}^{t} - \boldsymbol{w}_{k}) \right]$, we have
\begin{equation}\label{eq:per-main1}
\begin{aligned}
 &||\bar{\boldsymbol{w}}_k^{t+1} - \boldsymbol{w}^*||^2 = ||\bar{\boldsymbol{w}}_{k}^{t} - \eta \left[g_{k}^{t}\bar{\boldsymbol{x}}_{k}^t + \lambda (\bar{\boldsymbol{w}}_{k}^{t} - \boldsymbol{w}_{k}) \right] - \boldsymbol{w}^*||^2 
 \\&=||\bar{\boldsymbol{w}}_k^t - \boldsymbol{w}^*||^2 + \underbrace{\eta^2||g_{k}^{t}\bar{\boldsymbol{x}}_{k}^t + \lambda (\bar{\boldsymbol{w}}_{k}^{t} - \boldsymbol{w}_{k})||^2}_{(P1)} 
 \\&+ \underbrace{2\eta\langle \boldsymbol{w}^* - \bar{\boldsymbol{w}}_k^t, g_k^t \bar{\boldsymbol{x}}_k^t + \lambda(\bar{\boldsymbol{w}}_{k}^{t} - \boldsymbol{w}_{k})\rangle}_{(P2)}.
\end{aligned}
\end{equation}
To bound ($P1$), we have
\begin{equation} \label{equa:per-p1support}
\begin{aligned}
&||g_k^t \bar{\boldsymbol{x}}_k^t + \lambda(\bar{\boldsymbol{w}}_k^t -\boldsymbol{w}_k)||^2 \\
&= ||g_k^t \bar{\boldsymbol{x}}_k^t - \bar{g}_k^t \bar{\boldsymbol{x}}_k^t||^2 + \langle g_k^t \bar{\boldsymbol{x}}_k^t - \bar{g}_k^t \bar{\boldsymbol{x}}_k^t, \bar{g}_k^t \bar{\boldsymbol{x}}_k^t + \lambda(\bar{\boldsymbol{w}}_k^t -\boldsymbol{w}_k)    \rangle \\
&\hspace{1em} + ||\bar{g}_k^t \bar{\boldsymbol{x}}_k^t + \lambda(\bar{\boldsymbol{w}}_k^t -\boldsymbol{w}_k)||^2,
\end{aligned}
\end{equation}
where the second term is zero, and thus we focus on bounding the last term in~\ref{equa:per-p1support}. We have,
\begin{equation} \label{equa:per-p1support1}
\begin{aligned}
&||\bar{g}_k^t \bar{\boldsymbol{x}}_k^t + \lambda(\bar{\boldsymbol{w}}_k^t -\boldsymbol{w}_k)||^2 
\\&= ||\bar{g}_k^t \bar{\boldsymbol{x}}_k^t||^2 +\lambda^2||\bar{\boldsymbol{w}}_k^t -\boldsymbol{w}_k||^2 +2\lambda\langle \bar{g}_k^t \bar{\boldsymbol{x}}_k^t, \bar{\boldsymbol{w}}_k^t -\boldsymbol{w}_k    \rangle
\\& \leq ||\bar{g}_k^t \bar{\boldsymbol{x}}_k^t||^2 + 2\lambda^2||\bar{\boldsymbol{w}}_k^t -\boldsymbol{w}^*||^2 + 2\lambda^2||\boldsymbol{w}^* -\boldsymbol{w}_k ||^2 
\\&\hspace{7em}+2\lambda\frac{1}{N} \sum_{n\in\mathcal{N}} \langle \nabla F(\boldsymbol{w}_{n,k}^t) \bar{\boldsymbol{x}}_k^t, \bar{\boldsymbol{w}}_k^t -\boldsymbol{w}_k    \rangle,
\end{aligned}
\end{equation}
where we used the Cauchy–Schwarz inequality, i.e., $||a+b||^2 \leq 2||a||^2 +2||b||^2$ for any two real-valued vectors $a$ and $b$. To bound the last term in~\ref{equa:per-p1support1}, we have
\begin{equation} \label{equa:per-innerproduct-xandw}
\begin{aligned}
& \frac{1}{N} \sum_{n\in\mathcal{N}} \langle \nabla F(\boldsymbol{w}_{n,k}^t) \bar{\boldsymbol{x}}_k^t, \bar{\boldsymbol{w}}_k^t -\boldsymbol{w}_k    \rangle 
\\
&= \left(1+ \frac{\eta}{2}+\frac{1}{\eta} \right)\bar{\boldsymbol{x}}_k^t||\bar{g}_k^t||^2+\frac{1}{2\eta}||\boldsymbol{w}^*-\boldsymbol{w}_k||^2 
\\
&+(1+\eta)\frac{1}{N} \sum_{n\in\mathcal{N}}||\bar{\boldsymbol{w}}_k^t -\boldsymbol{w}_{n,k}^t||^2
\\
&+ \bar{\boldsymbol{x}}_k^t \frac{1}{N} \sum_{n\in\mathcal{N}}(F(\boldsymbol{w}_{n,k}^t) - F(\boldsymbol{w}^*)) + \bar{\boldsymbol{x}}_k^t \frac{L_p}{2}|| \bar{\boldsymbol{w}}_k^t- \boldsymbol{w}^*||^2,
\end{aligned}
\end{equation}
where we used the fact $||a+b||^2 \leq (1+z)||a||^2 +(1+1/z)||b||^2$ (for any two real valued vectors $a$ and $b$ and $z > 0$) and $L_p$-smoothness of $F(\boldsymbol{w}_{n,k}^t)$ in the second inequality and $\frac{1}{N} \sum_{n\in\mathcal{N}}||\boldsymbol{w}_{n,k}^t - \boldsymbol{w}^*||^2 = ||\bar{\boldsymbol{w}}_k^t - \boldsymbol{w}^*||^2$ in the last equality. By plugging~\ref{equa:per-innerproduct-xandw} into~\ref{equa:per-p1support1}, we have
\begin{equation} 
\begin{aligned}
&||\bar{g}_k^t \bar{\boldsymbol{x}}_k^t + \lambda(\bar{\boldsymbol{w}}_k^t -\boldsymbol{w}_k)||^2 \leq (2\lambda^2+\bar{\boldsymbol{x}}_k^t\lambda L_p)||\bar{\boldsymbol{w}}_k^t -w^*||^2 
\\&+ \bigg(1+2\lambda+\lambda\eta +\frac{2\lambda}{\eta} \bigg) \bar{\boldsymbol{x}}_k^t ||\bar{g}_k^t||^2 +(2 \lambda^2 + \frac{\lambda}{\eta}) ||\boldsymbol{w}^* - \boldsymbol{w}_k||^2
\\&+ 2\lambda(1+\eta)\frac{1}{N} \sum_{n\in\mathcal{N}}||\bar{\boldsymbol{w}}_k^t -\boldsymbol{w}_{n,k}^t||^2
\\&+ 2\lambda \bar{\boldsymbol{x}}_k^t \frac{1}{N} \sum_{n\in\mathcal{N}}(F(\boldsymbol{w}_{n,k}^t) - F(\boldsymbol{w}^*)) 
\end{aligned}
\end{equation}
Replacing the above result back in \ref{equa:per-p1support}, we obtain
\begin{equation} \label{equa:per-p1support-final}
\begin{aligned}
&||g_k^t \bar{\boldsymbol{x}}_k^t + \lambda(\bar{\boldsymbol{w}}_k^t -\boldsymbol{w}_k)||^2 
\\& \leq ||g_k^t \bar{\boldsymbol{x}}_k^t - \bar{g}_k^t \bar{\boldsymbol{x}}_k^t||^2 +(2\lambda^2+\bar{\boldsymbol{x}}_k^t\lambda L_p)||\bar{\boldsymbol{w}}_k^t -w^*||^2  
\\&+ \bigg(1+2\lambda+\lambda\eta +\frac{2\lambda}{\eta} \bigg) \bar{\boldsymbol{x}}_k^t ||\bar{g}_k^t||^2 +(2 \lambda^2 + \frac{\lambda}{\eta}) ||\boldsymbol{w}^* - \boldsymbol{w}_k||^2
\\&+ 2\lambda(1+\eta)\frac{1}{N} \sum_{n\in\mathcal{N}}||\bar{\boldsymbol{w}}_k^t -\boldsymbol{w}_{n,k}^t||^2
\\&+ 2\lambda \bar{\boldsymbol{x}}_k^t \frac{1}{N} \sum_{n\in\mathcal{N}}(F(\boldsymbol{w}_{n,k}^t) - F(\boldsymbol{w}^*)) 
\end{aligned}
\end{equation}
We now bound the term ($P2$). We have
\begin{equation} \label{equa:per-p1support-boundP2}
\begin{aligned}
&\langle \boldsymbol{w}^* - \bar{\boldsymbol{w}}_k^t, g_k^t \bar{\boldsymbol{x}}_k^t+ \lambda(\bar{\boldsymbol{w}}_k^t -\boldsymbol{w}_k)\rangle
\\&=\langle \boldsymbol{w}^* - \bar{\boldsymbol{w}}_k^t, g_k^t \bar{\boldsymbol{x}}_k^t -\bar{g}_k^t \bar{\boldsymbol{x}}_k^t \rangle + \langle \boldsymbol{w}^* - \bar{\boldsymbol{w}}_k^t, \bar{g}_k^t \bar{\boldsymbol{x}}_k^t \rangle 
\\&\hspace{14em}+ \lambda\langle \boldsymbol{w}^* - \bar{\boldsymbol{w}}_k^t, \bar{\boldsymbol{w}}_k^t -\boldsymbol{w}_k\rangle
\\
&\leq \eta|| \bar{\boldsymbol{w}}_k^t -\boldsymbol{w}^*||^2+\frac{1}{2\eta}||g_k^t \bar{\boldsymbol{x}}_k^t-\bar{g}_k^t \bar{\boldsymbol{x}}_k^t||^2 +\frac{1}{2\eta}||\bar{g}_k^t \bar{\boldsymbol{x}}_k^t||^2
\\&\hspace{10em}+ \lambda\langle \boldsymbol{w}^* - \bar{\boldsymbol{w}}_k^t, \bar{\boldsymbol{w}}_k^t -\boldsymbol{w}_k\rangle.
\end{aligned}
\end{equation}
For the last term, we have 
\begin{equation} 
\begin{aligned}
\langle \boldsymbol{w}^* - \bar{\boldsymbol{w}}_k^t, \bar{\boldsymbol{w}}_k^t -\boldsymbol{w}_k\rangle &\leq \frac{\eta}{2}||\bar{\boldsymbol{w}}_k^t-\boldsymbol{w}^*||^2 + \frac{1}{\eta}||\bar{\boldsymbol{w}}_k^t- \boldsymbol{w}^*||^2 
\\&+ \frac{1}{\eta}||\boldsymbol{w}_k - \boldsymbol{w}^*||^2.
\end{aligned}
\end{equation}
Replacing it in \ref{equa:per-p1support-boundP2}, we have
\begin{equation} 
\begin{aligned}
&\langle \boldsymbol{w}^* - \bar{\boldsymbol{w}}_k^t, g_k^t \bar{\boldsymbol{x}}_k^t+ \lambda(\bar{\boldsymbol{w}}_k^t -\boldsymbol{w}_k)\rangle
\leq \left(\eta + \frac{\lambda\eta}{2} + \frac{\lambda}{\eta}\right)||\bar{\boldsymbol{w}}_k^t-\boldsymbol{w}^*||^2
\\&\hspace{2em}+\frac{1}{2\eta}||g_k^t \bar{\boldsymbol{x}}_k^t-\bar{g}_k^t \bar{\boldsymbol{x}}_k^t||^2 +\frac{1}{2\eta}||\bar{g}_k^t \bar{\boldsymbol{x}}_k^t||^2 + \frac{\lambda}{\eta}||\boldsymbol{w}^* - \boldsymbol{w}_k||^2.
\end{aligned}
\end{equation}
Therefore, \ref{eq:per-main1} can be rewritten as
\begin{equation} \label{equa:per-p1support-bound_extra}
\begin{aligned}
 &||\bar{\boldsymbol{w}}_k^{t+1} - \boldsymbol{w}^*||^2 
 \\&\leq \left(1+\eta + \frac{\lambda \eta}{2} + \frac{\lambda}{\eta} + 2\lambda^2 + \bar{\boldsymbol{x}}_k^t \lambda L_p \right)||\bar{\boldsymbol{w}}_k^t - \boldsymbol{w}^*||^2 
 \\&+ (1+\frac{1}{2\eta})||g_k^t \bar{\boldsymbol{x}}_k^t-\bar{g}_k^t \bar{\boldsymbol{x}}_k^t||^2 
 \\&+ \left( 2\lambda + 2\lambda^2 \eta^2 + \lambda\eta \right) ||\boldsymbol{w}^* - \boldsymbol{w}_k||^2 
 \\&+ 2\lambda \eta^2 (1+\eta)\frac{1}{N} \sum_{n\in\mathcal{N}}||\bar{\boldsymbol{w}}_k^t -\boldsymbol{w}_{n,k}^t||^2
\\&+ \left(2\lambda \eta^2 \bar{\boldsymbol{x}}_k^t + 2L_p\left(1 + 2\lambda + \lambda \eta + \frac{2\lambda}{\eta} + \frac{1}{2\eta}\right)\bar{\boldsymbol{x}}_k^t\right) 
\\& \times \frac{1}{N} \sum_{n\in\mathcal{N}}(F(\boldsymbol{w}_{n,k}^t) - F(\boldsymbol{w}^*)) ,
\end{aligned}
\end{equation}
where we used the fact $||\bar{g}_k^t||^2 = \frac{1}{N} \sum_{n\in\mathcal{N}} ||\nabla F(\boldsymbol{w}_{n,k}^t)||^2 \leq 2L_p \frac{1}{N} \sum_{n\in\mathcal{N}}(F(\boldsymbol{w}_{n,k}^t) - F(\boldsymbol{w}^*))$. For the last term, we have
\begin{equation} 
\begin{aligned}
&\frac{1}{N} \sum_{n\in\mathcal{N}}(F(\boldsymbol{w}_{n,k}^t) - F(\boldsymbol{w}^*))
\\&= L_p\frac{1}{N} \sum_{n\in\mathcal{N}}||\bar{\boldsymbol{w}}_{k}^t-\boldsymbol{w}_{n,k}^t||^2
+ \left(\frac{1}{2L_p}+\frac{1}{2\vartheta}\right) ||\nabla F(\bar{\boldsymbol{w}}_{k}^t)||^2,
\end{aligned}
\end{equation}
where we applied the $L_p$-smoothness of $F_n(.)$ and the Polyak-Lojasiewicz condition in the first inequality. By replacing the above result back into \ref{equa:per-p1support-bound_extra} and taking the expectation, we obtain
\begin{equation} \label{equa:per-p1support-bound_final}
\begin{aligned}
&\mathbb{E}||\bar{\boldsymbol{w}}_k^{t+1} - \boldsymbol{w}^*||^2 
 \\&\leq \left(1+\eta + \frac{\lambda \eta}{2} + \frac{\lambda}{\eta} + 2\lambda^2 + \bar{\boldsymbol{x}}_k^t \lambda L_p \right)\mathbb{E}||\bar{\boldsymbol{w}}_k^t - \boldsymbol{w}^*||^2 
 \\&+ (1+\frac{1}{2\eta})\mathbb{E}||g_k^t \bar{\boldsymbol{x}}_k^t-\bar{g}_k^t \bar{\boldsymbol{x}}_k^t||^2 
 \\&+ \left( 2\lambda + 2\lambda^2 \eta^2 + \lambda\eta \right) \mathbb{E}||\boldsymbol{w}^* - \boldsymbol{w}_k||^2 
 \\&+ \bigg(2\lambda \eta^2 (1+\eta) + \bigg(2\lambda \eta^2 \bar{\boldsymbol{x}}_k^t 
 \\&+ 2L_p\left(1 + 2\lambda + \lambda \eta + \frac{2\lambda}{\eta} + \frac{1}{2\eta}\right)\bar{\boldsymbol{x}}_k^t\bigg) \bigg)
 \\& \times \mathbb{E}\bigg(\frac{1}{N} \sum_{n\in\mathcal{N}}||\bar{\boldsymbol{w}}_k^t -\boldsymbol{w}_{n,k}^t||^2\bigg)
\\&+ \left(2\lambda \eta^2 \bar{\boldsymbol{x}}_k^t + 2L_p\left(1 + 2\lambda + \lambda \eta + \frac{2\lambda}{\eta} + \frac{1}{2\eta}\right)\bar{\boldsymbol{x}}_k^t\right) 
\\& \times \left(\frac{1}{2L_p}+\frac{1}{2\vartheta}\right) ||\nabla F(\bar{\boldsymbol{w}}_{k}^t)||^2.
\end{aligned}
\end{equation}

\subsection{Proof of Theorem \ref{Theorem1}} \label{ProofTheorem1}

\noindent
From Lemmas \ref{Lemma1}, \ref{Lemma2}, and \ref{Lemma3}, we have
\begin{equation} \label{equa:updaterule_final}
\begin{aligned}
&\mathbb{E}||\bar{\boldsymbol{w}}_k^{t+1} - \boldsymbol{w}^*||^2 
 \\&\leq \left(1+\eta + \frac{\lambda \eta}{2} + \frac{\lambda}{\eta} + 2\lambda^2 + \bar{\boldsymbol{x}}_k^t \lambda L_p \right)\mathbb{E}||\bar{\boldsymbol{w}}_k^t - \boldsymbol{w}^*||^2  
 \\&+ (1+\frac{1}{2\eta})|\bar{\boldsymbol{x}}_k^t|^2 \sum_{n \in \mathcal{N}}\frac{\nu N_{h} D Tr(H^{2})}{2MN^2} 
 \\&+ \left( 2\lambda + 2\lambda^2 \eta^2 + \lambda\eta \right) \mathbb{E}||\boldsymbol{w}^* - \boldsymbol{w}_k||^2 
 \\&+ \frac{1}{2}\eta \bigg(2\lambda \eta^2 (1+\eta) 
 \\&+ \bigg(2\lambda \eta^2 \bar{\boldsymbol{x}}_k^t + 2L_p\left(1 + 2\lambda + \lambda \eta + \frac{2\lambda}{\eta} + \frac{1}{2\eta}\bigg)\bar{\boldsymbol{x}}_k^t\right) \bigg)T G^2
\\&+ \left(2\lambda \eta^2 \bar{\boldsymbol{x}}_k^t + 2L_p\left(1 + 2\lambda + \lambda \eta + \frac{2\lambda}{\eta} + \frac{1}{2\eta}\right)\bar{\boldsymbol{x}}_k^t\right) 
\\& \hspace{7em} \times \left(\frac{1}{2L_p}+\frac{1}{2\vartheta}\right) ||\nabla F(\bar{\boldsymbol{w}}_{k}^t)||^2.
\end{aligned}
\end{equation}
Let us define $Y_k^t = \mathbb{E}||\bar{\boldsymbol{w}}_k^t - \boldsymbol{w}^*||^2$
and $\Phi_k = \frac{1}{2}\eta \bigg(2\lambda \eta^2 (1+\eta) + \bigg(2\lambda \eta^2 \bar{\boldsymbol{x}}_k^t + 2L_p\left(1 + 2\lambda + \lambda \eta + \frac{2\lambda}{\eta} + \frac{1}{2\eta}\bigg)\bar{\boldsymbol{x}}_k^t\right) \bigg)T G^2 + (1+\frac{1}{2\eta})|\bar{\boldsymbol{x}}_k^t|^2 \sum_{n \in \mathcal{N}}\frac{\nu N_{h} D Tr(H^{2})}{2MN^2}$. From \ref{equa:updaterule_final}, we have $\sum_{t=1}^{T}Y_k^{t+1} \leq \sum_{t=0}^{T-1} \left(1+\eta + \frac{\lambda \eta}{2} + \frac{\lambda}{\eta} + 2\lambda^2 + \bar{\boldsymbol{x}}_k^t \lambda L_p \right)Y_k^t + \Phi_k$.

\begin{table*}[!ht]
    \centering
    \caption{Difference in performance in quantum learning with the different number of layers across various datasets.}
    \label{tab:layer}
    \begin{tabular}{|c|cc|cc|cc|cc|}
    \toprule
    \multicolumn{1}{|c|}{\multirow{2}{*}{$l$ (num layers)}} & \multicolumn{2}{c|}{\textbf{MNIST}} & \multicolumn{2}{|c|}{\textbf{FashionMNIST}} & \multicolumn{2}{c|}{\textbf{CIFAR-100}} & \multicolumn{2}{c|}{\textbf{Caltech-101}} \\
    \cmidrule{2-9}
    & \textbf{Loss Value} & \textbf{Acc.} & \textbf{Loss Value} & \textbf{Acc.} & \textbf{Loss Value} & \textbf{Acc.} & \textbf{Loss Value} & \textbf{Acc.} \\
    \midrule
    1 & 0.1732 & 96.20\% & \textbf{0.2724} & \textbf{91.11\%} & \textbf{1.3388} & \textbf{55.63\%}  &  0.6747 & 87.75\% \\
    2 & \textbf{0.1612} & \textbf{97.13\%} & 0.2841 & 90.48\% & 1.3716 & 53.18\% & 0.6774 & 86.55\% \\
    3 & 0.2009 & 95.01\% & 0.2991 & 89.18\% & 1.3660 & 52.98\% & \textbf{0.5806} & \textbf{89.36\%}  \\
    4 & 0.1950 & 95.32\% & 0.3397 & 88.17\% & 1.4045 & 52.02\% & 0.7749 & 85.75\% \\
    5 & 0.3319 & 91.05\% & 0.3864 & 86.08\% & 1.4935 & 48.34\% & 0.7261 & 85.67\% \\
    10 & 0.4211 & 86.38\% & 0.5421 & 80.30\% & 1.8187 & 36.89\% & 1.1392 & 72.46\% \\
    \bottomrule
    \end{tabular}  
\end{table*}
\begin{table*}[!ht]
    \centering
    \footnotesize
    \caption{Difference in performance in quantum learning with the different number of qubits across various datasets.}
    \label{tab:qubit}
    \begin{tabular}{|c|cc|cc|cc|cc|}
    \toprule
    \multicolumn{1}{|c|}{\multirow{2}{*}{N (Num Qubit)}} & \multicolumn{2}{c|}{\textbf{MNIST}} & \multicolumn{2}{|c|}{\textbf{FashionMNIST}} & \multicolumn{2}{c|}{\textbf{CIFAR-100}} & \multicolumn{2}{c|}{\textbf{Caltech-101}} \\
    \cmidrule{2-9}
    & \textbf{Loss Value} & \textbf{Acc.} & \textbf{Loss Value} & \textbf{Acc.} & \textbf{Loss Value} & \textbf{Acc.} & \textbf{Loss Value} & \textbf{Acc.} \\
    \midrule
    1 & 2.311 & 9.88\% & 2.32 & 9.83\% & X & X & X & X \\
    2 & 2.108 & 20.68\% & 2.212 & 15.93\% & 4.026 & 11.54\% & 2.266 & 14.58\% \\
    5 & 1.872 & 43.33\% & 1.667 & 33.87\% & 3.011 & 29.78\% & 2.051 & 26.78\% \\
    10 & \textbf{0.109} & \textbf{97.13\%} & \textbf{0.283} & \textbf{90.33\%} & \textbf{1.320} & \textbf{55.04\%} & \textbf{1.366} & \textbf{52.98\%} \\
    \bottomrule
    \end{tabular}
\end{table*}

\noindent
We define a diminishing stepsize $\eta = \frac{4\theta}{k+\omega}$ for some $\theta >\frac{1}{4\mu}$ and $\omega >0$. By defining $m =\max \{\frac{\theta^2\Phi_k}{4\theta\mu-1}, (\omega+1)Y_0\}$, we prove that $Y_k \leq \frac{m}{k+\omega}$ by induction. Due to $4\theta\mu >1$, we have $Y_{k+1} = \left(1+\eta + \frac{\lambda \eta}{2} + \frac{\lambda}{\eta} + 2\lambda^2 + \bar{\boldsymbol{x}}_k^t \lambda L_p \right)Y_k^t + \Phi_k \leq \bigg( \frac{m}{k+\omega} + \frac{4\theta}{(k+\omega)^2}m + \frac{\lambda}{2} \frac{4\theta}{(k+\omega)^2}m + \frac{2\lambda^2}{k+\omega}m + \frac{\bar{\boldsymbol{x}}_k^t \lambda L_p}{k+\omega}m \bigg)+ \Phi_k$.
Assume we choose $\theta = \frac{3}{\mu}$ and $\omega = \frac{L}{\mu}$ , it follows that $m =\max \{\frac{\theta^2\Phi_k}{3\theta\mu-1}, (\omega+1)Y_0\} \leq \frac{\theta^2\Phi_k}{3\theta\mu-1} + (\omega+1)Y_0 = \frac{9\Phi_k}{8\mu^2} + \left( \frac{L}{\mu}+1\right)Y_0$.Therefore, $\mathbb{E}\left[F(\bar{\boldsymbol{w}}_k)\right] -F^* \leq \frac{L}{2}Y_k\leq \frac{L}{2}\frac{m}{(k+\omega)} \leq \frac{L}{2(k+L/\mu)} \left[\frac{9\Phi_k}{8\mu^2} + \left( \frac{L}{\mu}+1\right) \mathbb{E}||\boldsymbol{w}_0 - \boldsymbol{w}^*||^2 \right]$. Finally, by applying \ref{equa:final_convergenceIIDnew} recursively, the convergence bound of SPQFL after $K$ global communication rounds can be given as $\mathbb{E}\left[F(\boldsymbol{w}_K)\right] -F^* \leq \frac{L}{2(K+L/\mu)}
\left[\frac{9\Phi_K}{8\mu^2} + \left( \frac{L}{\mu}+1\right) \mathbb{E}||\boldsymbol{w}_0 - \boldsymbol{w}^*||^2 \right]$, where $\Phi_K = \frac{1}{2}\eta \bigg(2\lambda \eta^2 (1+\eta) + \bigg(2\lambda \eta^2 \bar{\boldsymbol{x}}_K + 2L_p\left(1 + 2\lambda + \lambda \eta + \frac{2\lambda}{\eta} + \frac{1}{2\eta}\right)\bar{\boldsymbol{x}}_K\bigg) \bigg)T G^2 + (1+\frac{1}{2\eta})|\bar{\boldsymbol{x}}_K|^2 \sum_{n \in \mathcal{N}}\frac{\nu N_{h} D Tr(H^{2})}{2MN^2}$, which completes the proof.

\section{Simulations and Evaluation} \label{Sec: Experiments}
\subsection{Simulation Settings} 
We consider a \textit{SPQFL} framework consisting of a single quantum server and 10 NISQ devices. Each quantum device trains a QNN model, i.e., image classification. 
We use the \textit{torchquantum} library for quantum simulations. The training procedure uses classical optimization approaches to update gate parameters, as well as gradient-based methods that take into account the system's quantum nature. This arrangement allows for the use of quantum mechanical features like superposition and entanglement in order to increase computational power and efficiency for specific types of data and operations.
\textcolor{black}{The QNN ansatz utilizes a hardware-efficient design where each circuit employs $n=4$ qubits started in the $\ket{0}^{\otimes n}$ state. It comprises of alternating layers of parameterized single-qubit rotations $R_Y(\theta_i)$ and $R_Z(\phi_i)$, followed by entangling \texttt{CZ} gates between surrounding qubits. Final measurements are made on the first qubit using a Pauli-$Z$ observable for classification. This approach efficiently balances expressibility and trainability while remaining compatible with existing NISQ hardware.}

\textbf{Quantum noise.} Quantum noise is created by integrating readout errors into quantum circuit measurement. This noise is very important since it directly influences the results of quantum computing, revealing errors in detecting qubit states, which is a very common scenario in quantum networks. In the implementation, each qubit is assigned a 3\% chance of inverting the measurement output. This configuration means that even if a quantum state is accurately prepared, the final readout may report the opposite state, giving a dimension of realism to the simulation by mirroring the flaws identified in the quantum measurement procedures utilized.
\textcolor{black}{To improve realism and address device-specific variability, we also use hardware-calibrated gate and gate noise profiles. 
 Single-qubit and two-qubit gate error probabilities, as well as qubit-dependent measurement errors, are extracted from genuine calibration data released by IBM Quantum and Rigetti devices. 
 These defects are fed into the simulation via the \texttt{torchquantum} noise model interface, assuring that each qubit and gate operation represents non-uniform, device-specific noise properties. 
 This improvement enables the simulation to more realistically replicate the stochastic behavior of NISQ hardware rather than depending on a uniform noise assumption.}
 
\begin{table*}[!ht]
\centering
\caption{\footnotesize Performance comparison between 3 different loss functions: CrossEntropy Loss, MSE Loss, and BCE Loss in all datasets.}
\label{table: comp loss}
\begin{tabular}{|p{2cm}|c|c|c|c|c|c|c|}
\hline
 Datasets & \multicolumn{2}{c|}{CrossEntropy Loss} & \multicolumn{2}{c|}{MSE Loss} & \multicolumn{2}{c|}{BCE Loss} \\
 \cline{2-7}
&Training Accuracy & Testing Accuracy & Training Accuracy & Testing Accuracy & Training Accuracy & Testing Accuracy \\
 \hline
MNIST & \textbf{96.21\%} & \textbf{94.93\%} & 94.67\% & 92.73\% & 95.32\% & 93.50\% \\
\hline
FashionMNIST & 90.71\% & \textbf{86.93\%} & 89.13\% & 85.22\% & \textbf{91.01\%} & 85.85\% \\
\hline
CIFAR-100 & \textbf{54.02\%} & \textbf{50.67\%} & 47.32\% & 44.39\% & 54.02\% & 48.13\% \\
\hline
Caltech-101 & \textbf{86.12\%} & \textbf{82.93\%} & 70.15\% & 65.53\% & 79.97\% & 75.13\% \\
\hline
\end{tabular}
\end{table*}
\begin{figure*}[ht!]
    \centering
    \footnotesize
    \begin{subfigure}[t]{0.245\linewidth} 
        \centering
        \includegraphics[width=\linewidth]{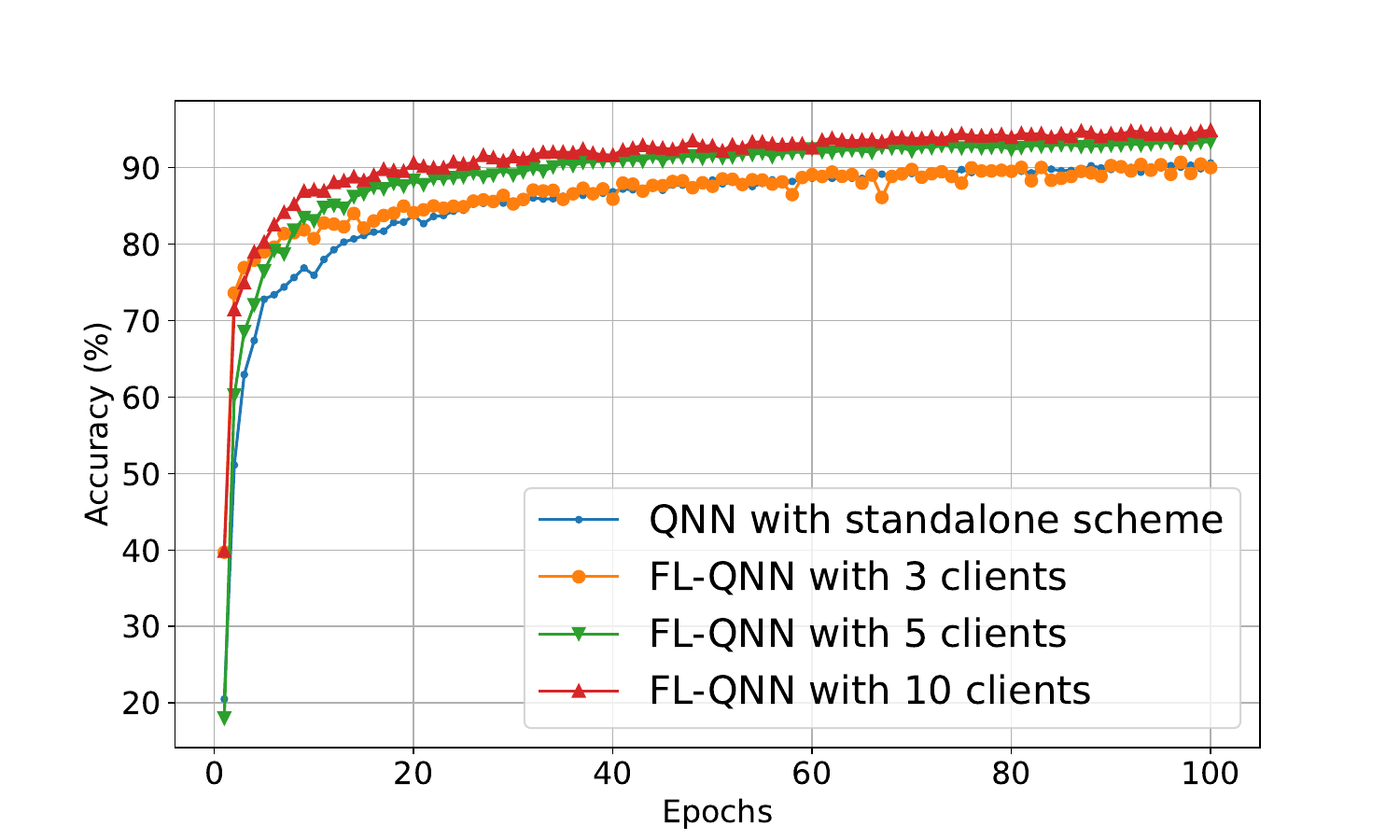}
        \caption{\footnotesize MNIST accuracy.}
    \end{subfigure}
    \hfill 
    \begin{subfigure}[t]{0.245\linewidth} 
        \centering
        \includegraphics[width=\linewidth]{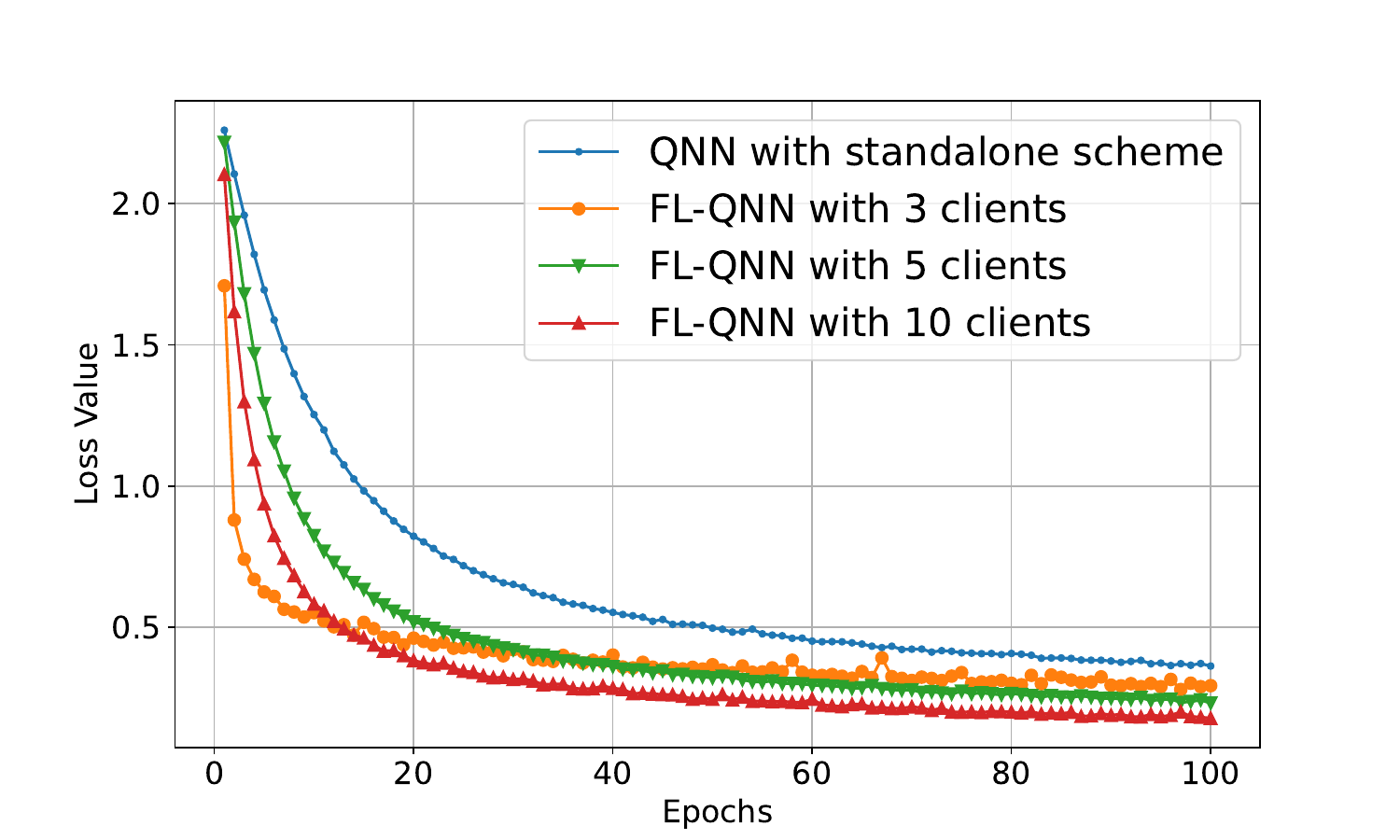}
        \caption{\footnotesize MNIST loss.}
    \end{subfigure}
    \begin{subfigure}[t]{0.245\linewidth} 
        \centering
        \includegraphics[width=\linewidth]{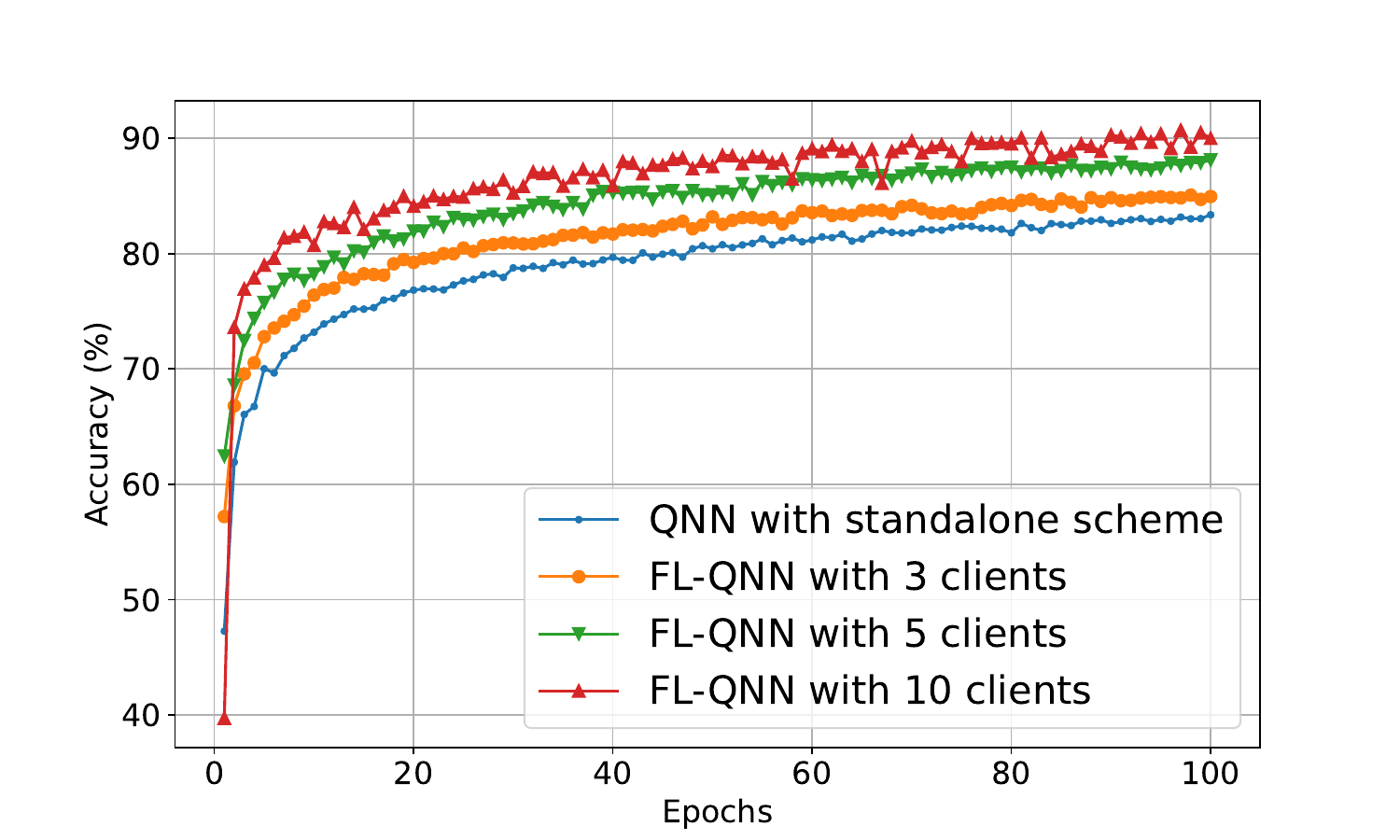}
        \caption{\footnotesize FashionMNIST accuracy.}
    \end{subfigure}
    \hfill 
    \begin{subfigure}[t]{0.245\linewidth} 
        \centering
        \includegraphics[width=\linewidth]{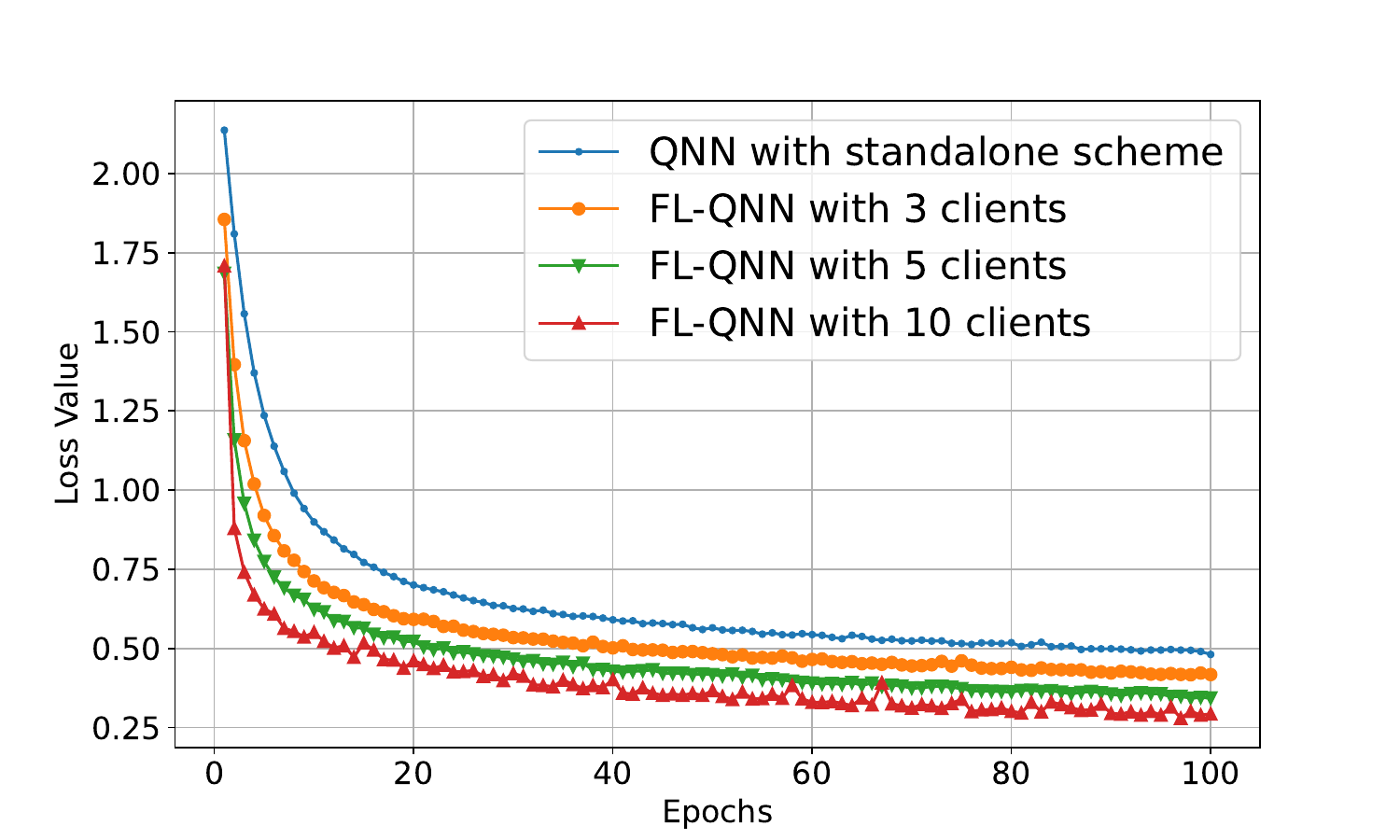}
        \caption{\footnotesize FashionMNIST loss.}
    \end{subfigure}
    \begin{subfigure}[t]{0.245\linewidth} 
        \centering
        \includegraphics[width=\linewidth]{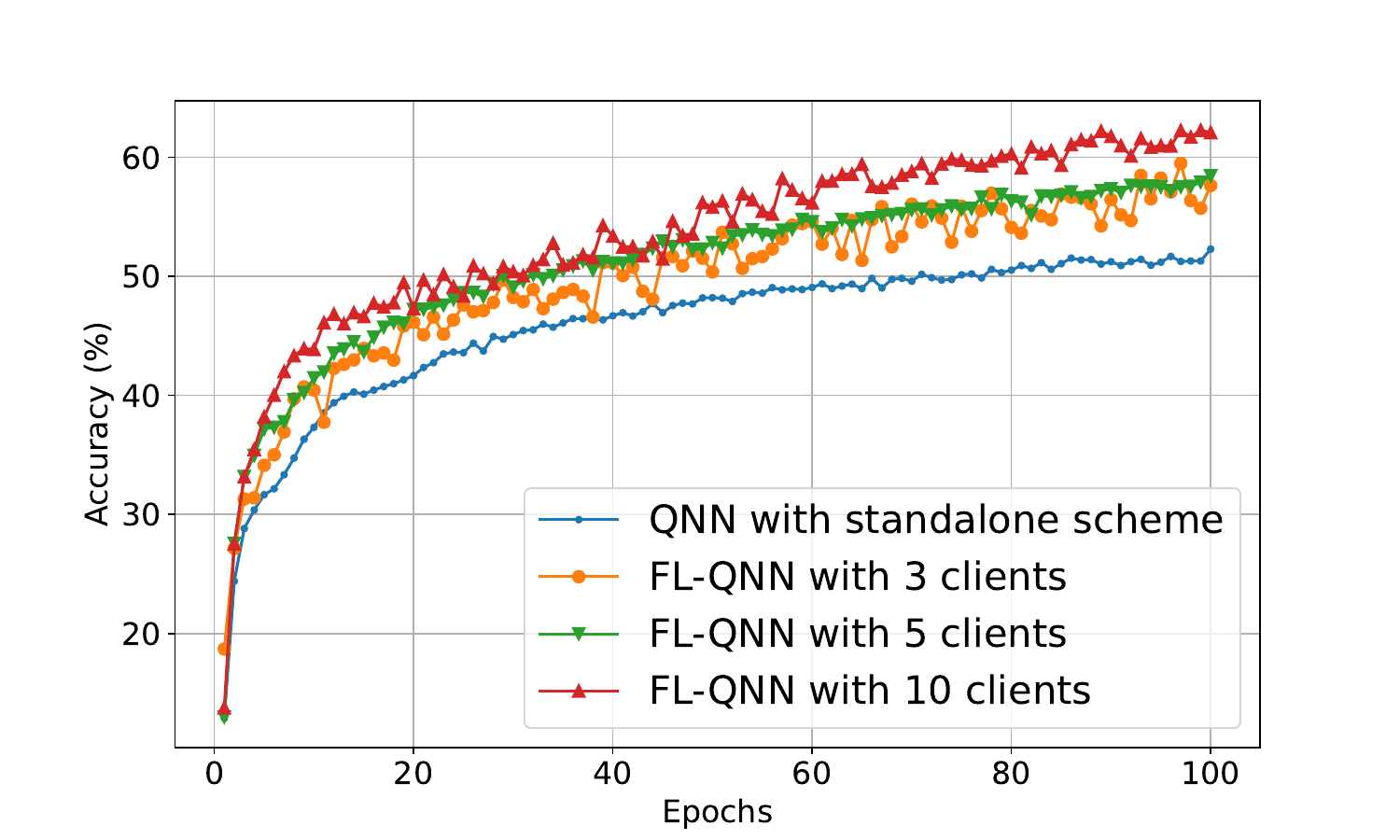}
        \caption{\footnotesize CIFAR-100 accuracy.}
    \end{subfigure}
    \hfill 
    \begin{subfigure}[t]{0.245\linewidth} 
        \centering
        \includegraphics[width=\linewidth]{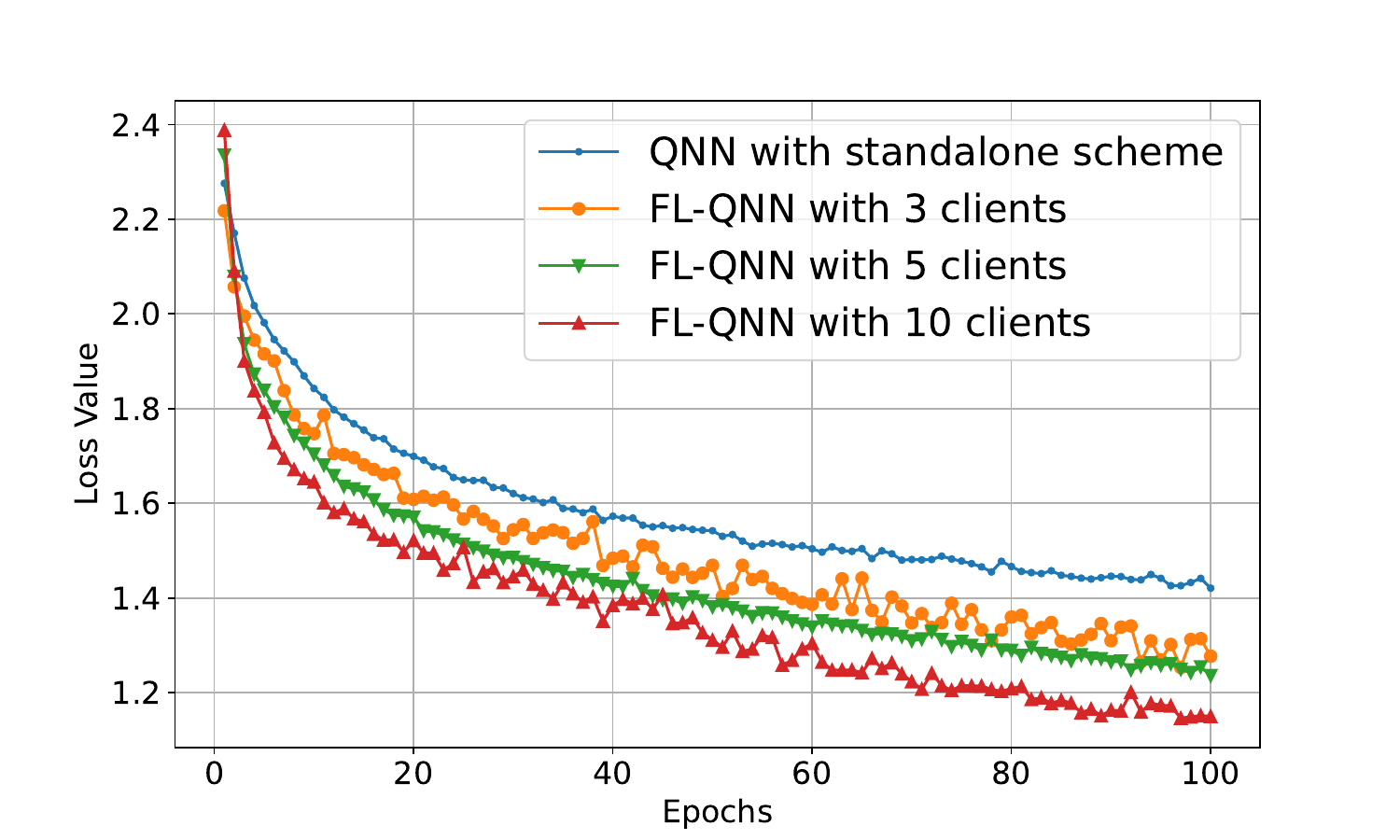}
        \caption{\footnotesize CIFAR-100 loss.}
    \end{subfigure}
    \begin{subfigure}[t]{0.245\linewidth} 
        \centering
        \includegraphics[width=\linewidth]{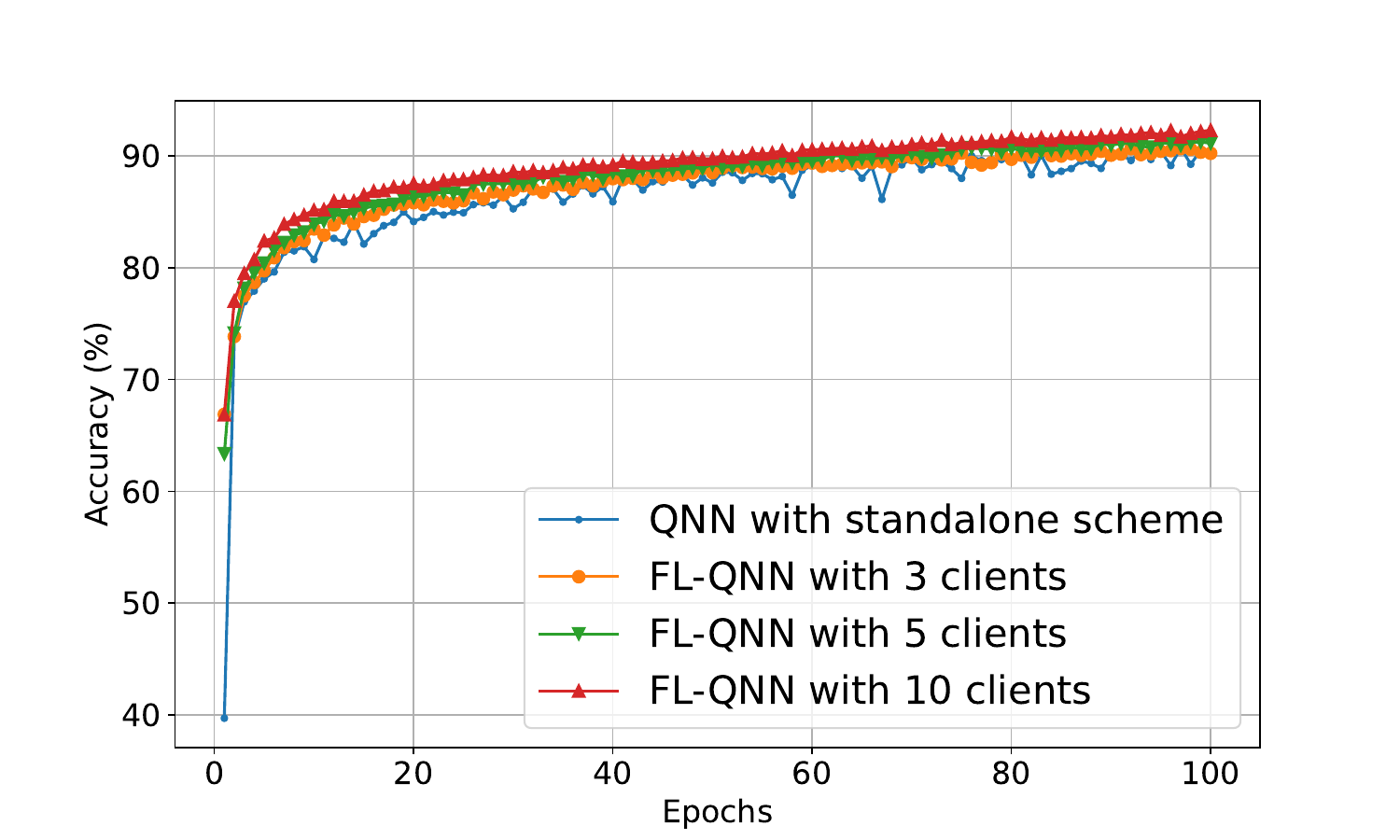}
        \caption{\footnotesize Caltech-101 accuracy.}
    \end{subfigure}
    \hfill 
    \begin{subfigure}[t]{0.245\linewidth} 
        \centering
        \includegraphics[width=\linewidth]{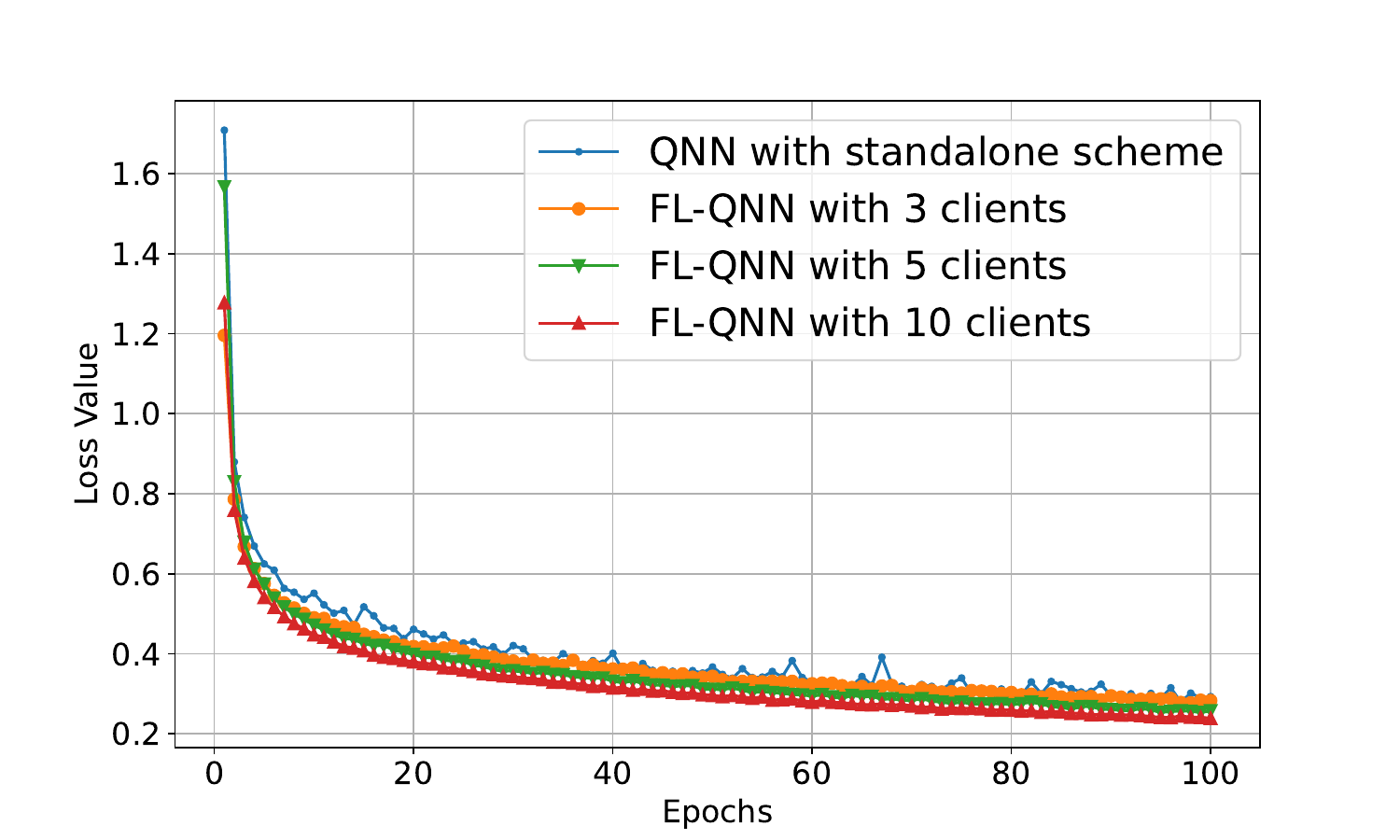}
        \caption{\footnotesize Caltech-101 loss.}
    \end{subfigure}
    \caption{\footnotesize Difference in performance in quantum learning with basic QNN and QFL Non-IID with 3,5, and 10 devices.}
    \label{fig: fl_niid}
\end{figure*}
\begin{figure*}[ht!]
    \centering
    \footnotesize
    \begin{subfigure}[t]{0.245\linewidth} 
        \centering
        \includegraphics[width=\linewidth]{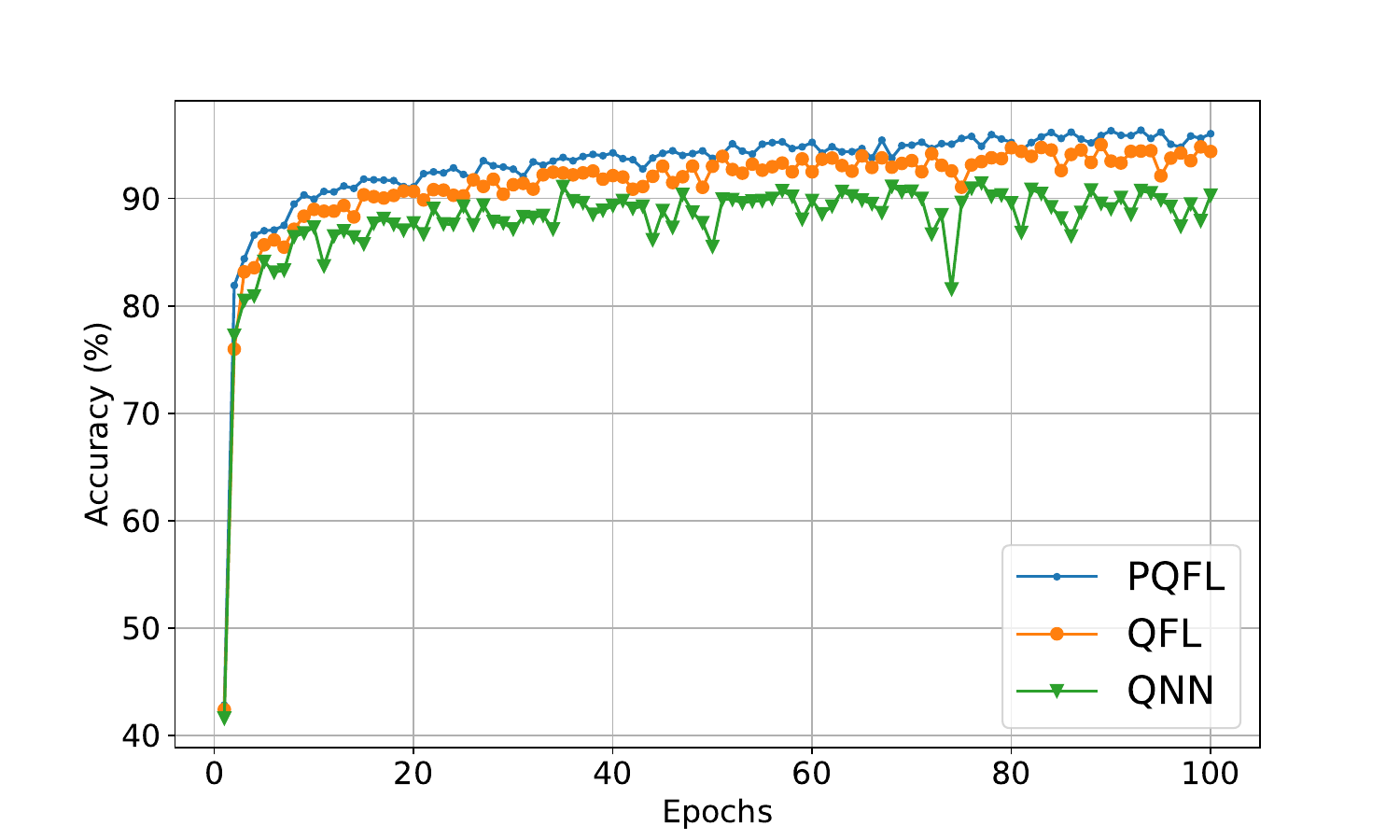}
        \caption{\footnotesize MNIST accuracy.}
    \end{subfigure}
    \hfill 
    \begin{subfigure}[t]{0.245\linewidth} 
        \centering
        \includegraphics[width=\linewidth]{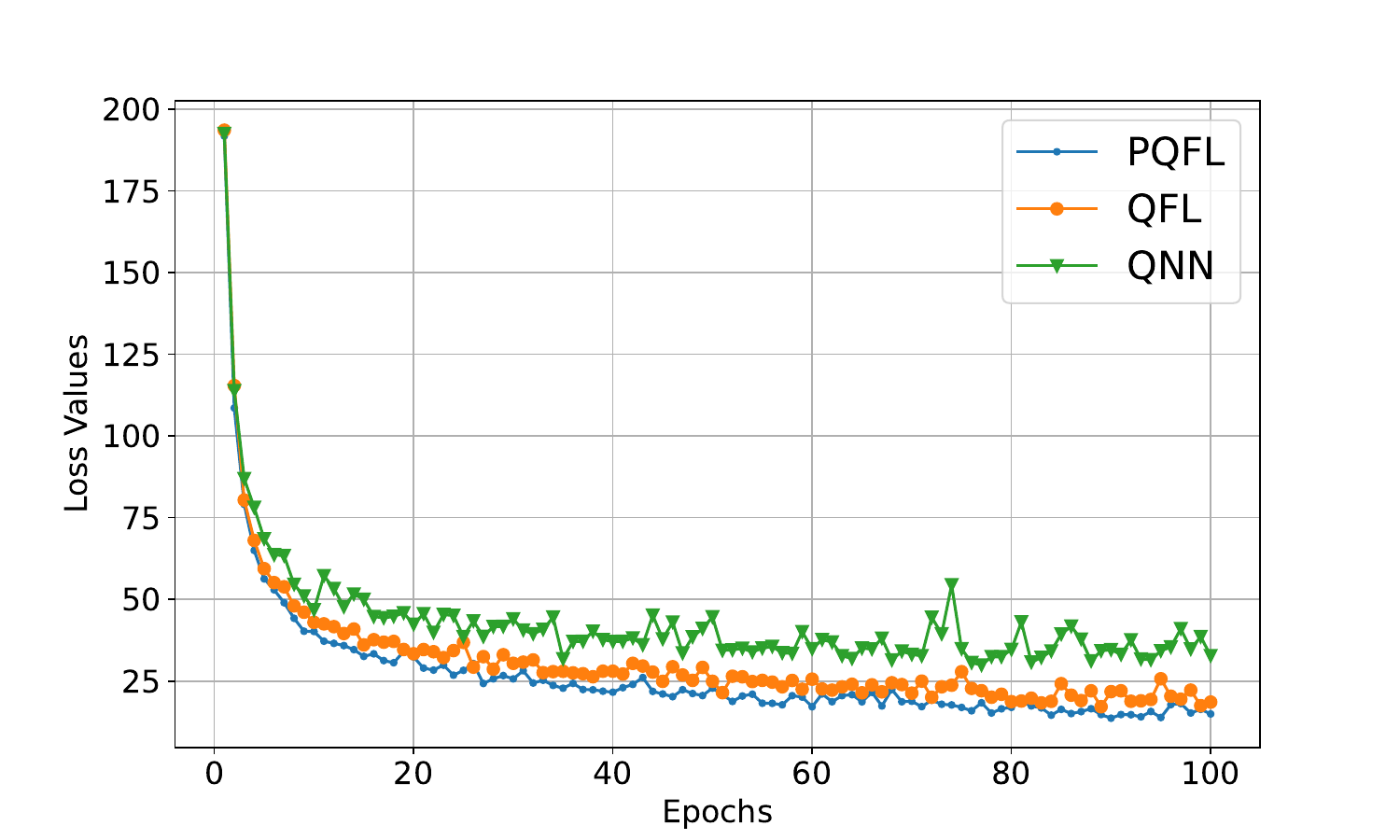}
        \caption{\footnotesize MNIST loss.}
    \end{subfigure}
    \begin{subfigure}[t]{0.245\linewidth} 
        \centering
        \includegraphics[width=\linewidth]{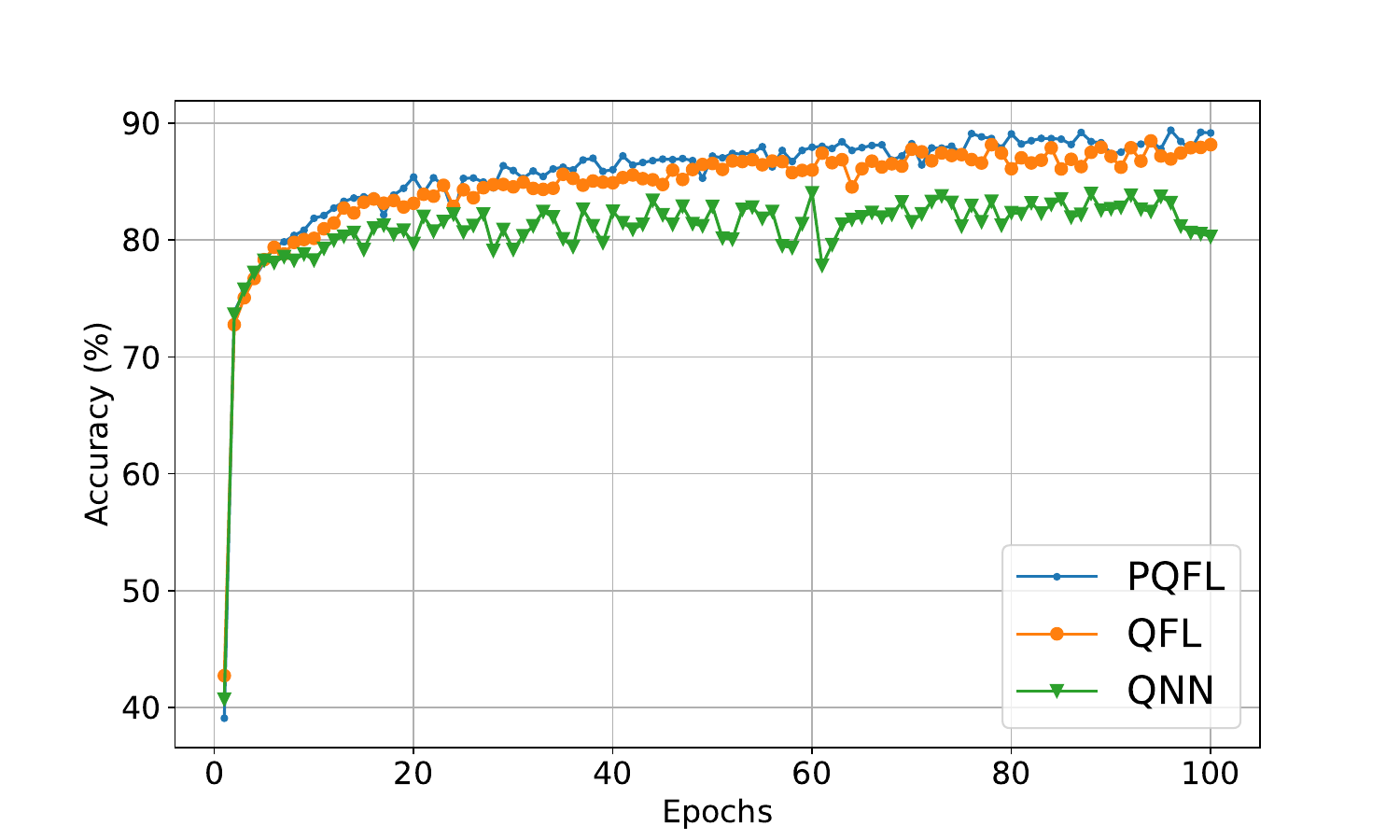}
        \caption{\footnotesize FashionMNIST accuracy.}
    \end{subfigure}
    \hfill 
    \begin{subfigure}[t]{0.245\linewidth} 
        \centering
        \includegraphics[width=\linewidth]{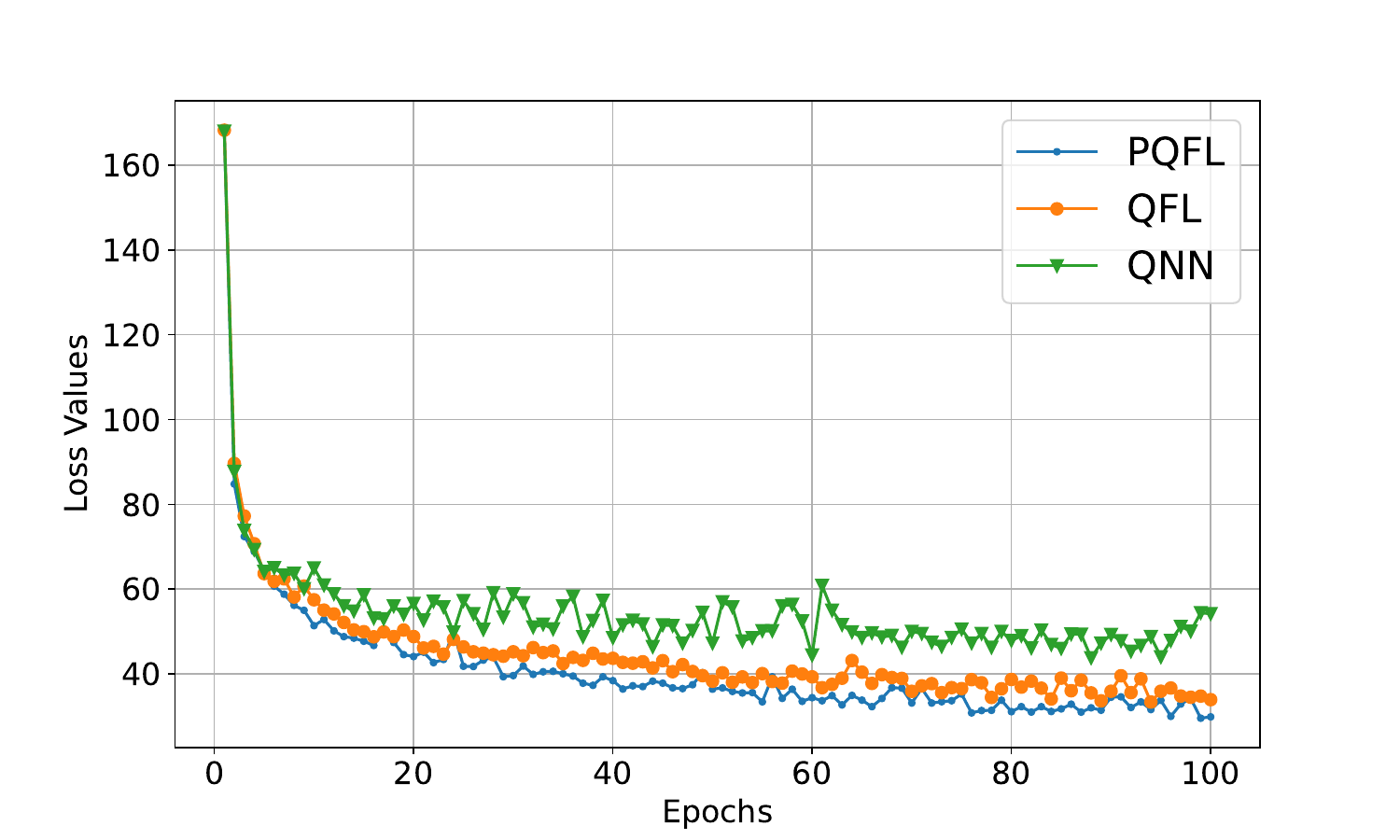}
        \caption{\footnotesize FashionMNIST loss.}
    \end{subfigure}
    \begin{subfigure}[t]{0.245\linewidth} 
        \centering
        \includegraphics[width=\linewidth]{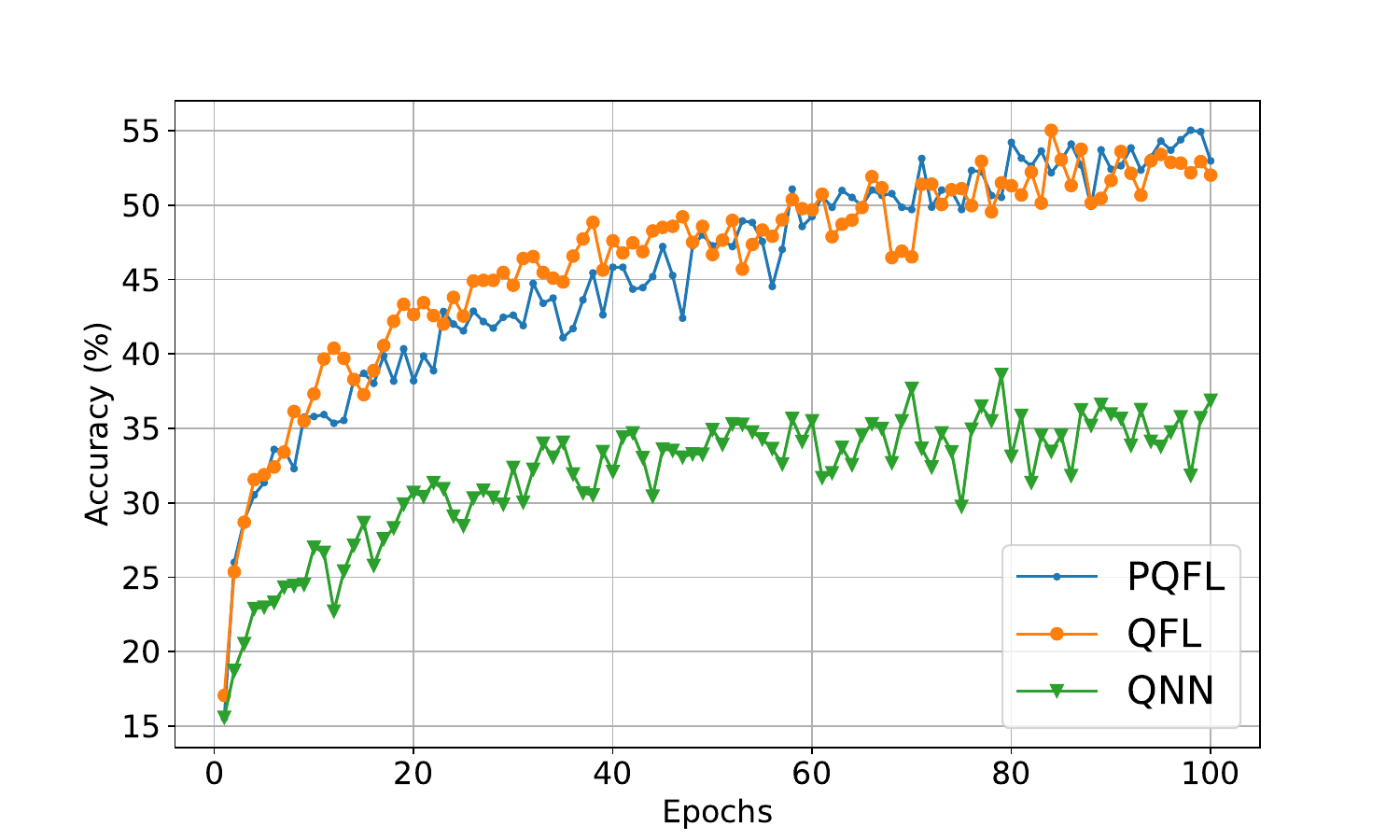}
        \caption{\footnotesize CIFAR-100 accuracy.}
    \end{subfigure}
    \hfill 
    \begin{subfigure}[t]{0.245\linewidth} 
        \centering
        \includegraphics[width=\linewidth]{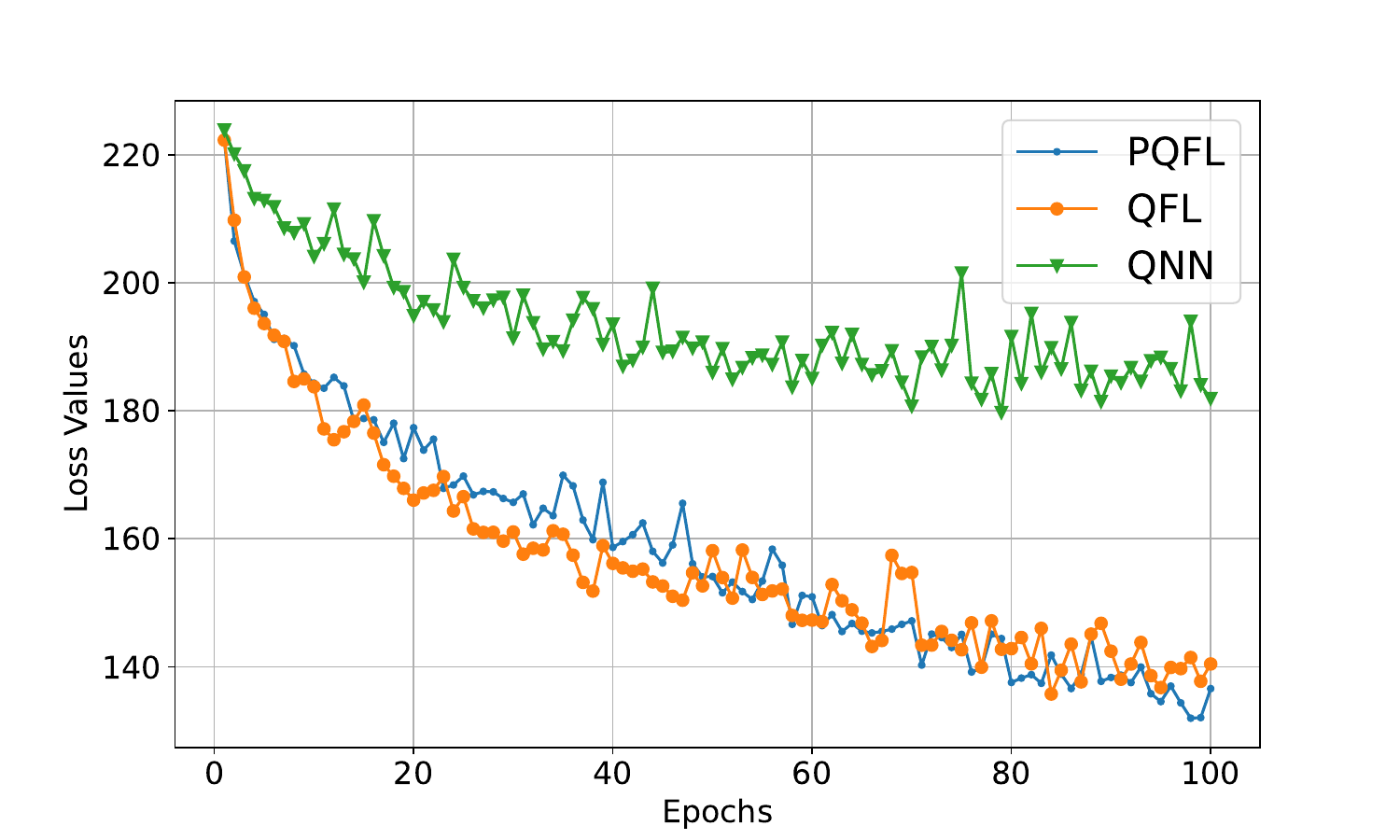}
        \caption{\footnotesize CIFAR-100 loss.}
    \end{subfigure}
    \begin{subfigure}[t]{0.245\linewidth} 
        \centering
        \includegraphics[width=\linewidth]{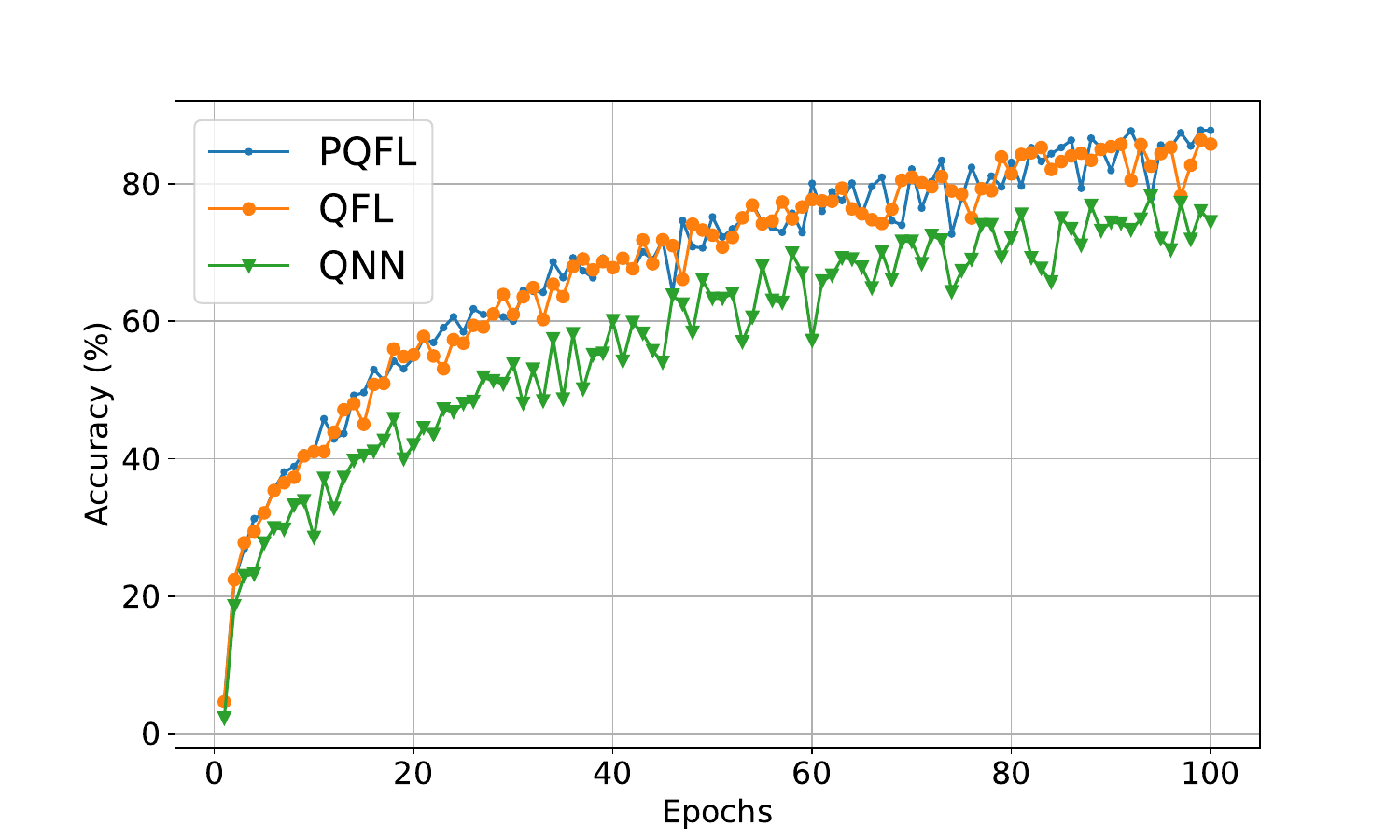}
        \caption{\footnotesize Caltech-101 accuracy.}
    \end{subfigure}
    \hfill 
    \begin{subfigure}[t]{0.245\linewidth} 
        \centering
        \includegraphics[width=\linewidth]{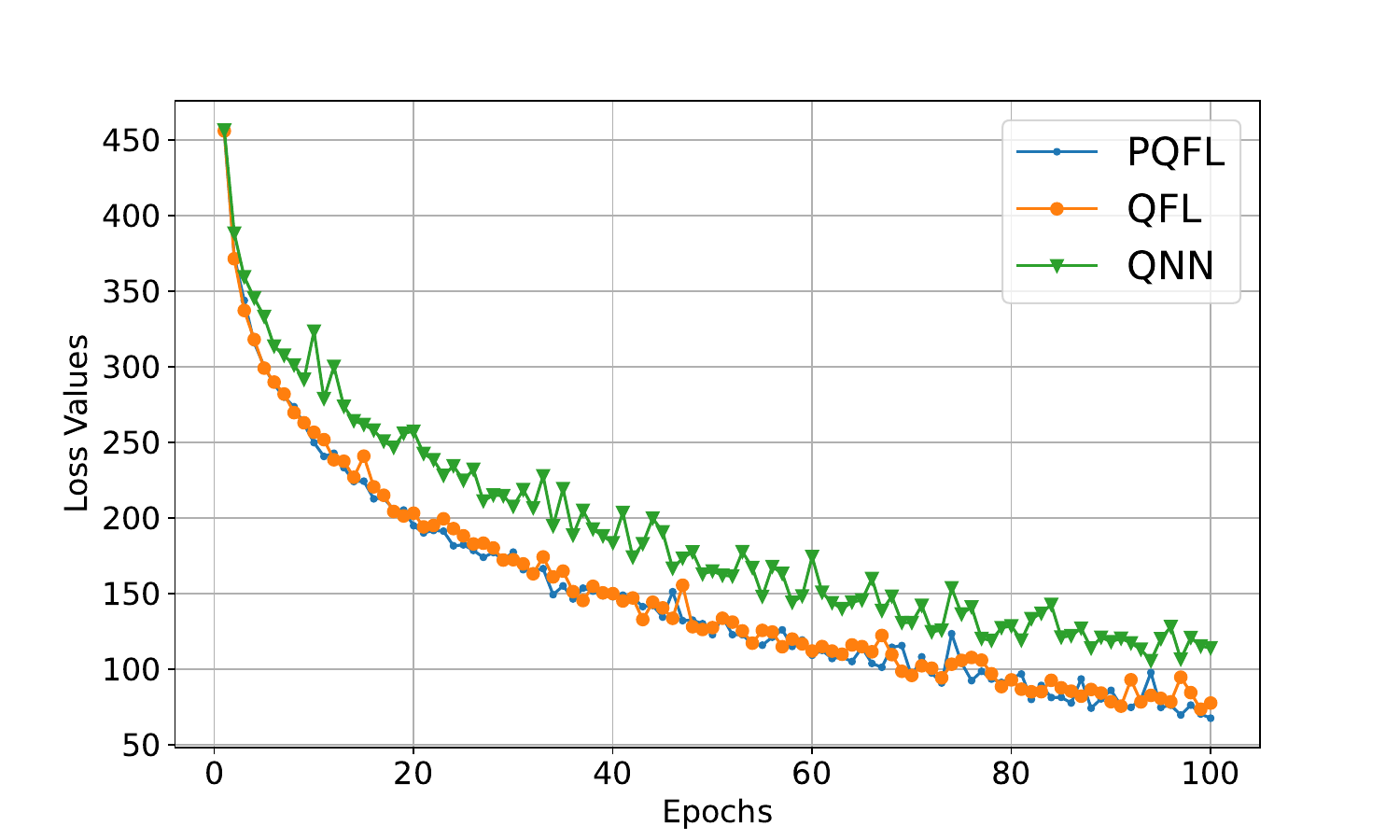}
        \caption{\footnotesize Caltech-101 loss.}
    \end{subfigure}
    \caption{\footnotesize Difference in performance in quantum learning with basic QNN, QFL, and PQFL for 10 devices. }
    \label{fig: fl_comp}
\end{figure*}

\textbf{Dataset.} To evaluate our framework's performance, we have run simulations on four established datasets: MNIST \cite{deng2012mnist}, FashionMNIST \cite{xiao2017fashion}, CIFAR-100 \cite{krizhevsky2009learning}, and Caltech-101 \cite{fei2004learning}. These datasets are chosen because they have a wide range of image types and complexity levels, making them an excellent benchmark for assessing classification algorithms. MNIST provides a basic dataset of grayscale handwritten digits, which is ideal for initial model testing. FashionMNIST's variety of clothing images creates a more difficult scenario, testing the model's ability to handle subtle class variations. CIFAR-100 broadens the scope by incorporating colored images from 100 categories, presenting the model with complex visual data. Caltech-101 tests the robustness of the model with images of varying scale, angle, and lighting conditions, resulting in more realistic scenarios. \textcolor{black}{Every experiment in this study involves a multi-class categorization challenge. For datasets like CIFAR-10, MNIST, and FashionMNIST, all classes were included in the training and assessment stages, with no subset selection. Every class $k \in \mathcal{K}$ corresponds to a unique observable $O_k$ (usually a Pauli-$Z$ operator measured on a specialized readout qubit), and the vector of expectation values $\mathbf{f}_n(\omega) = [f_{n,1}(\omega), f_{n,2}(\omega), \dots, f_{n,K}(\omega)]$ yields the final prediction.} Together, these datasets allow for a thorough evaluation of algorithmic performance across multiple contexts. All the simulations were performed on a server configured with an NVIDIA GeForce RTX 4090 GPU, 64GB of RAM, and Ubuntu 22.04 as the operating system.

{\color{black} \textbf{Non-Data distribution.} For our experiments only non-IID data distributions are considered as IID distributions does not consider heterogeneous data distributions and noises.} In settings with three quantum devices, the servers manage 25\%, 35\%, and 40\% of data, respectively. For configurations with five quantum devices, the data distribution is as follows: 14\%, 18\%, 22\%, 26\%, and 30\%. Furthermore, in an environment with ten quantum devices, data allocations range incrementally from 9\% to 16\%. As a result, the model is non-IID due to the unequal and diverse distribution of data across different devices, with each device containing distinct subsets of the overall dataset.

\begin{figure*}[ht!]
    \centering
    \footnotesize
    \begin{subfigure}[t]{0.245\linewidth} 
        \centering
        \includegraphics[width=\linewidth]{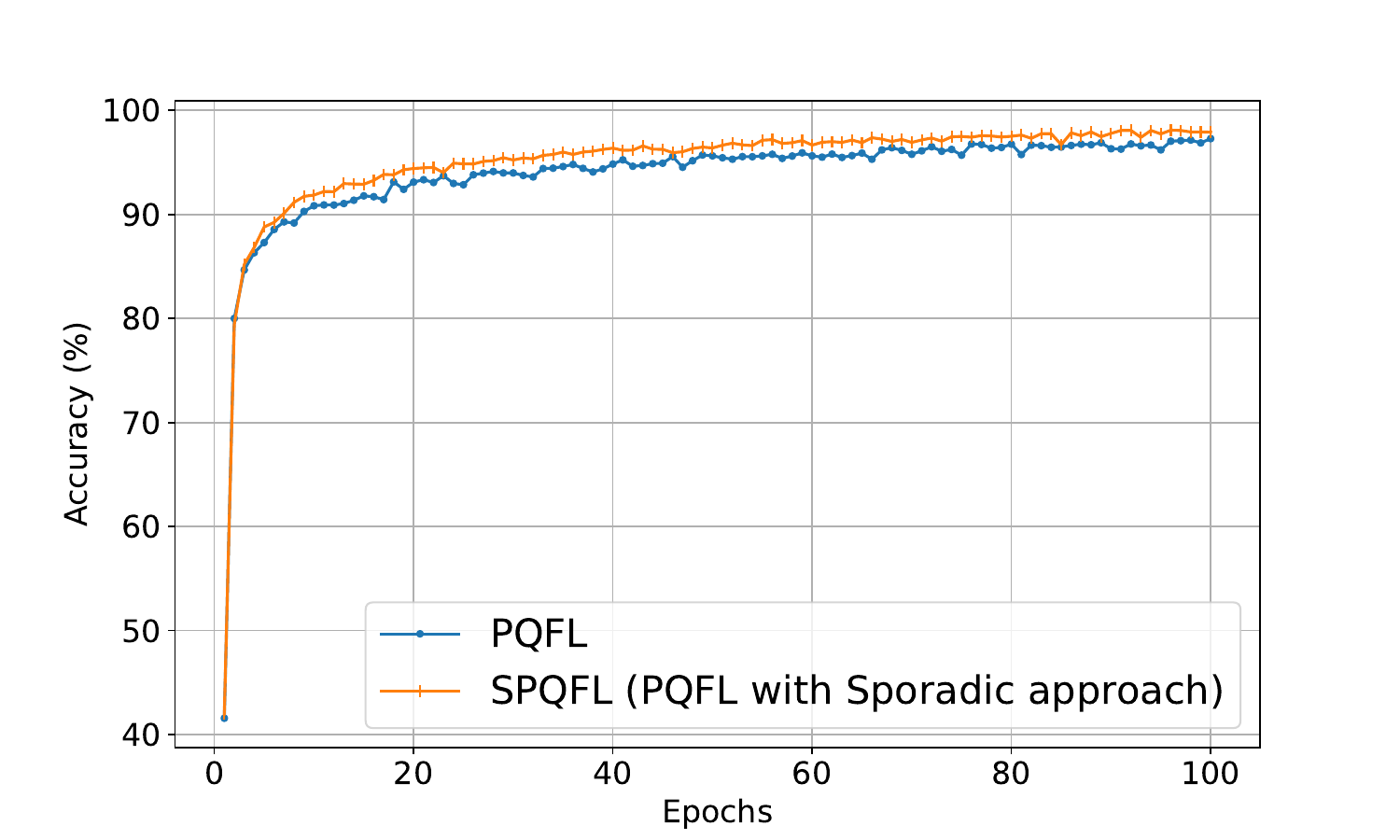}
        \caption{\footnotesize MNIST accuracy.}
    \end{subfigure}
    \hfill 
    \begin{subfigure}[t]{0.245\linewidth} 
        \centering
        \includegraphics[width=\linewidth]{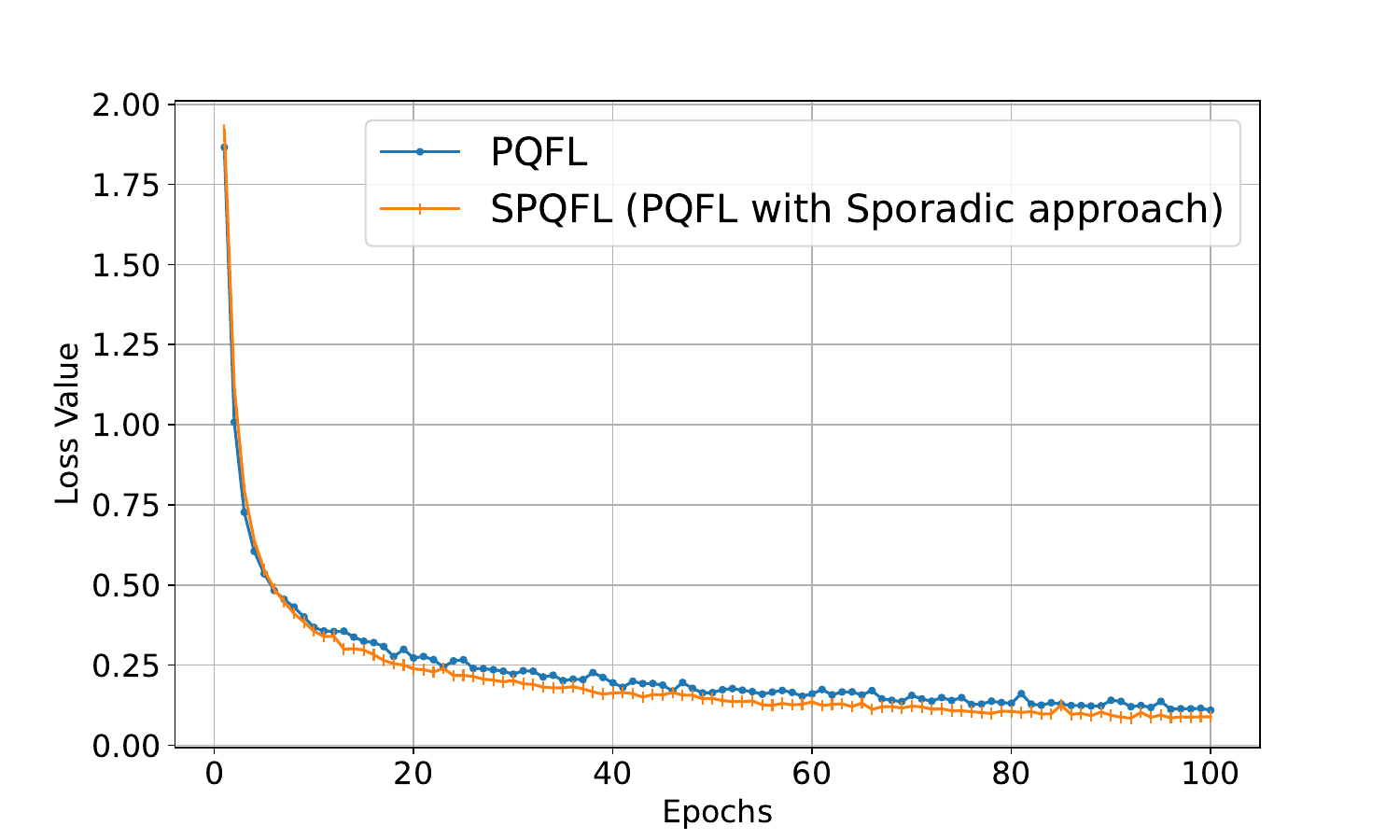}
        \caption{\footnotesize MNIST loss.}
    \end{subfigure}
    \begin{subfigure}[t]{0.245\linewidth} 
        \centering
        \includegraphics[width=\linewidth]{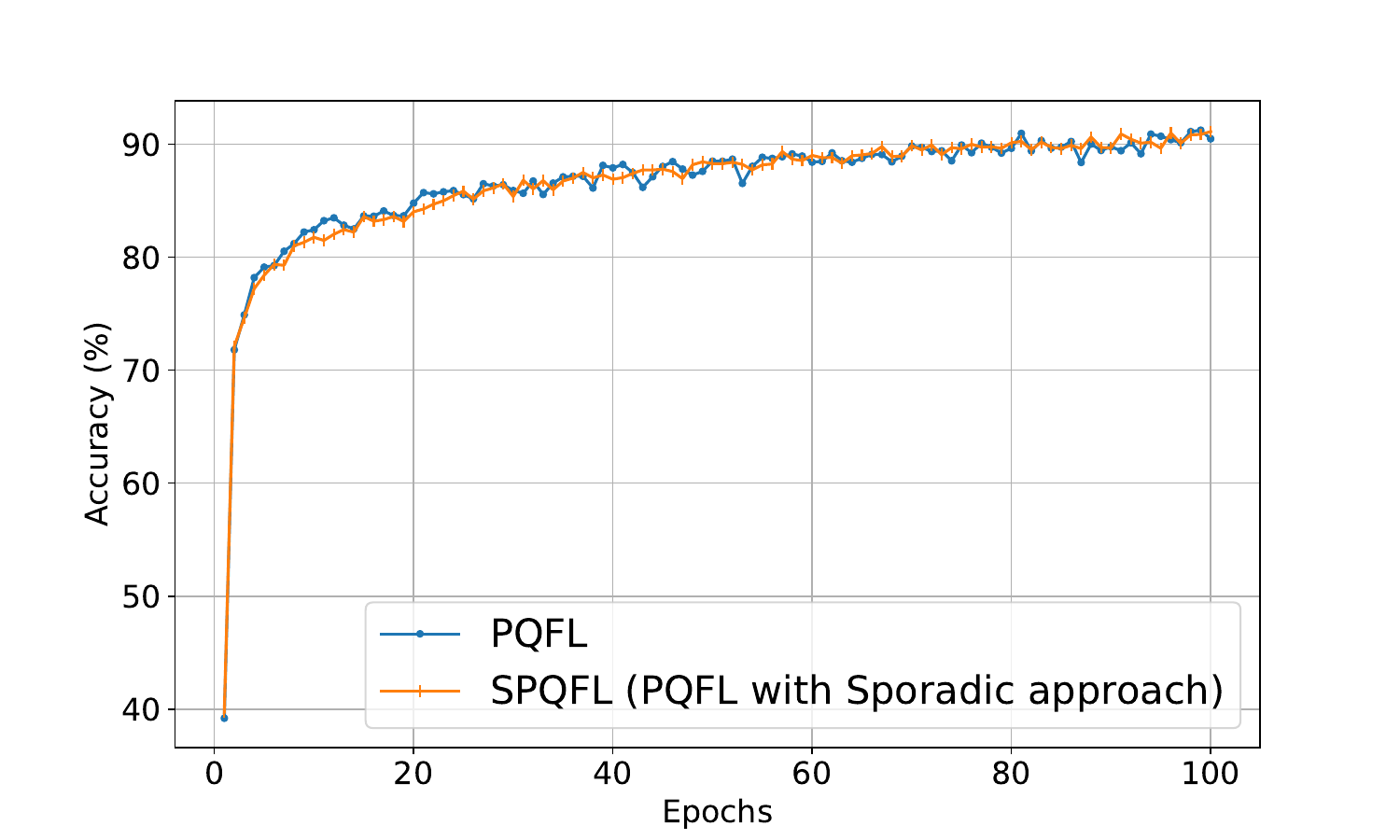}
        \caption{\footnotesize FashionMNIST accuracy.}
    \end{subfigure}
    \hfill 
    \begin{subfigure}[t]{0.245\linewidth} 
        \centering
        \includegraphics[width=\linewidth]{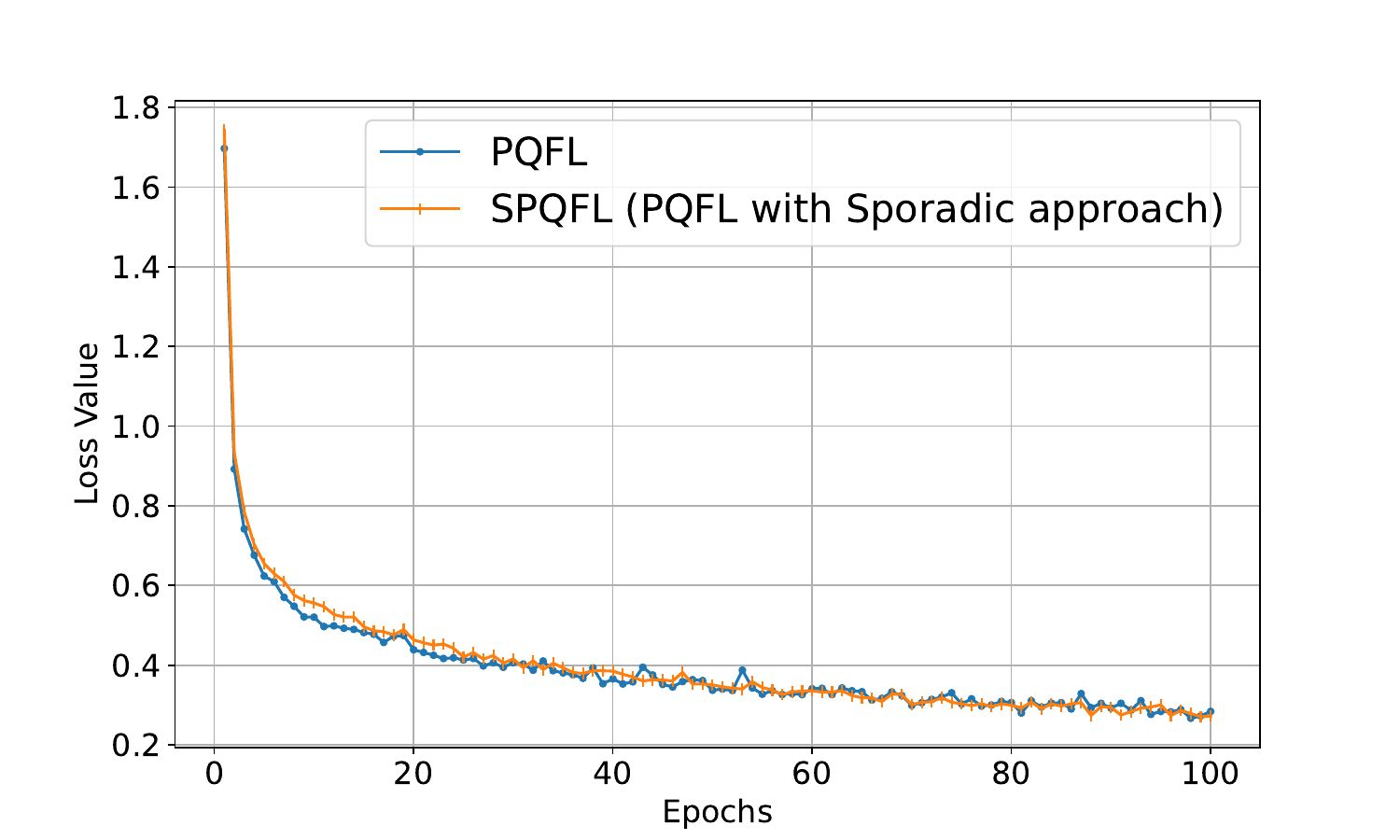}
        \caption{\footnotesize FashionMNIST loss.}
    \end{subfigure}
    \begin{subfigure}[t]{0.245\linewidth} 
        \centering
        \includegraphics[width=\linewidth]{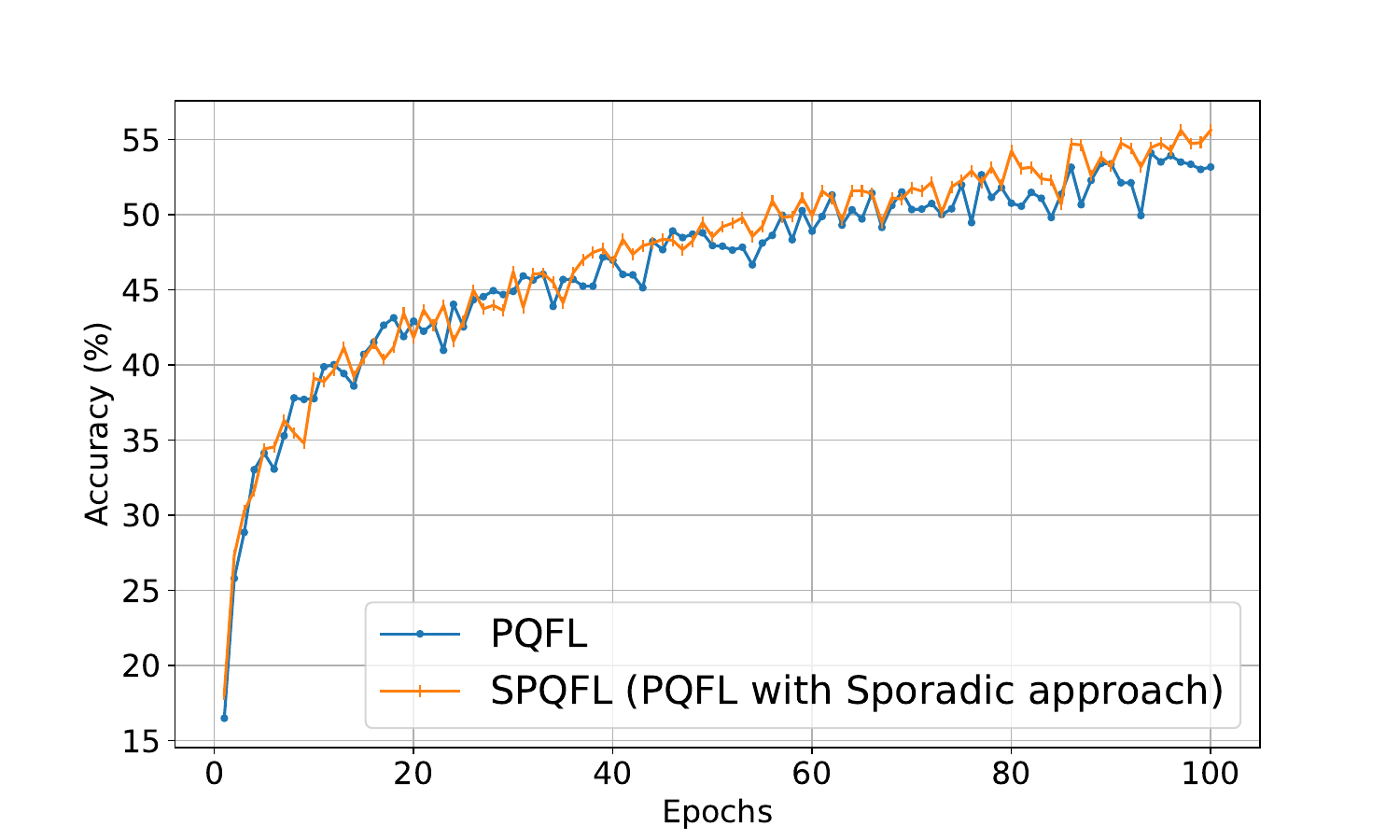}
        \caption{\footnotesize CIFAR-100 accuracy.}
    \end{subfigure}
    \hfill 
    \begin{subfigure}[t]{0.245\linewidth} 
        \centering
        \includegraphics[width=\linewidth]{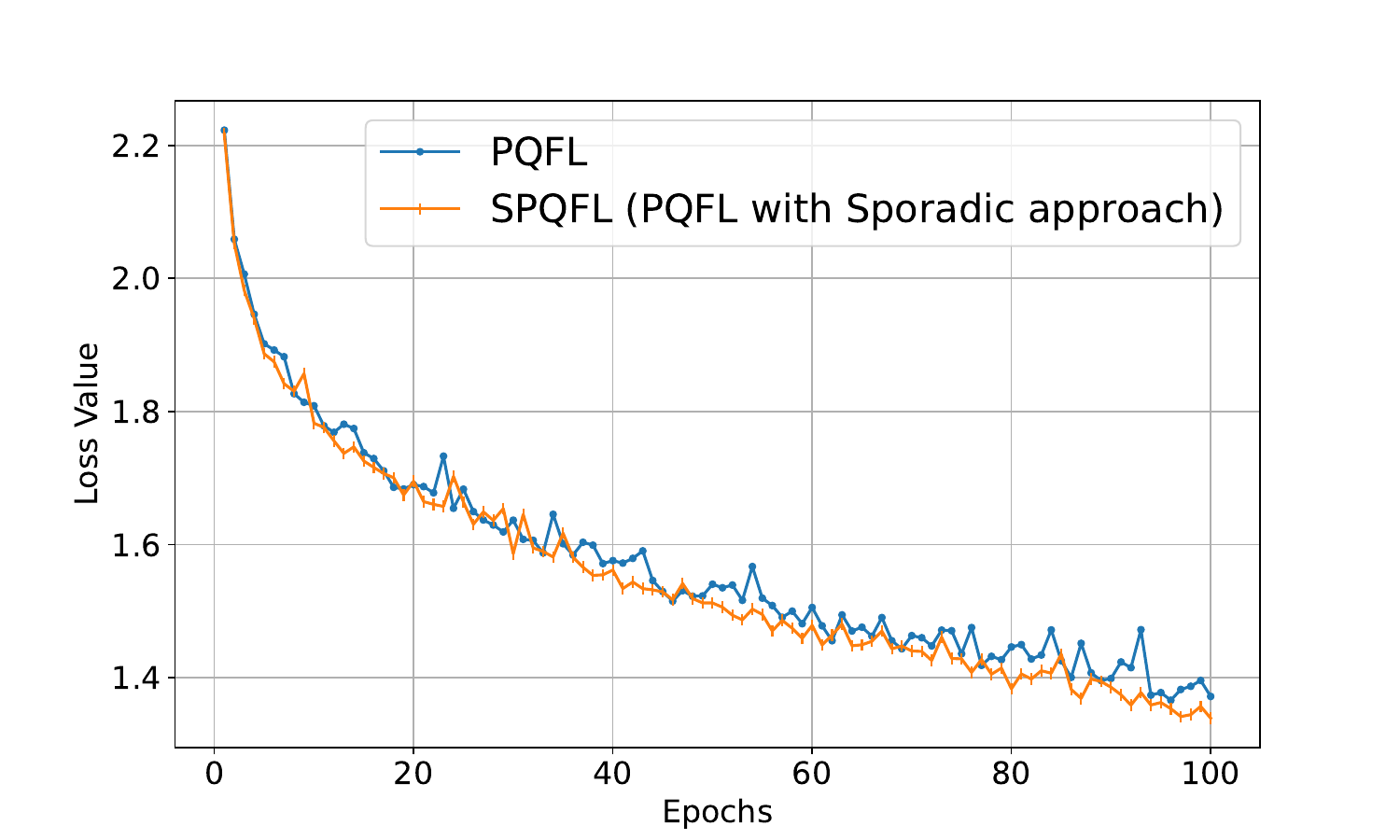}
        \caption{\footnotesize CIFAR-100 loss.}
    \end{subfigure}
    \begin{subfigure}[t]{0.245\linewidth} 
        \centering
        \includegraphics[width=\linewidth]{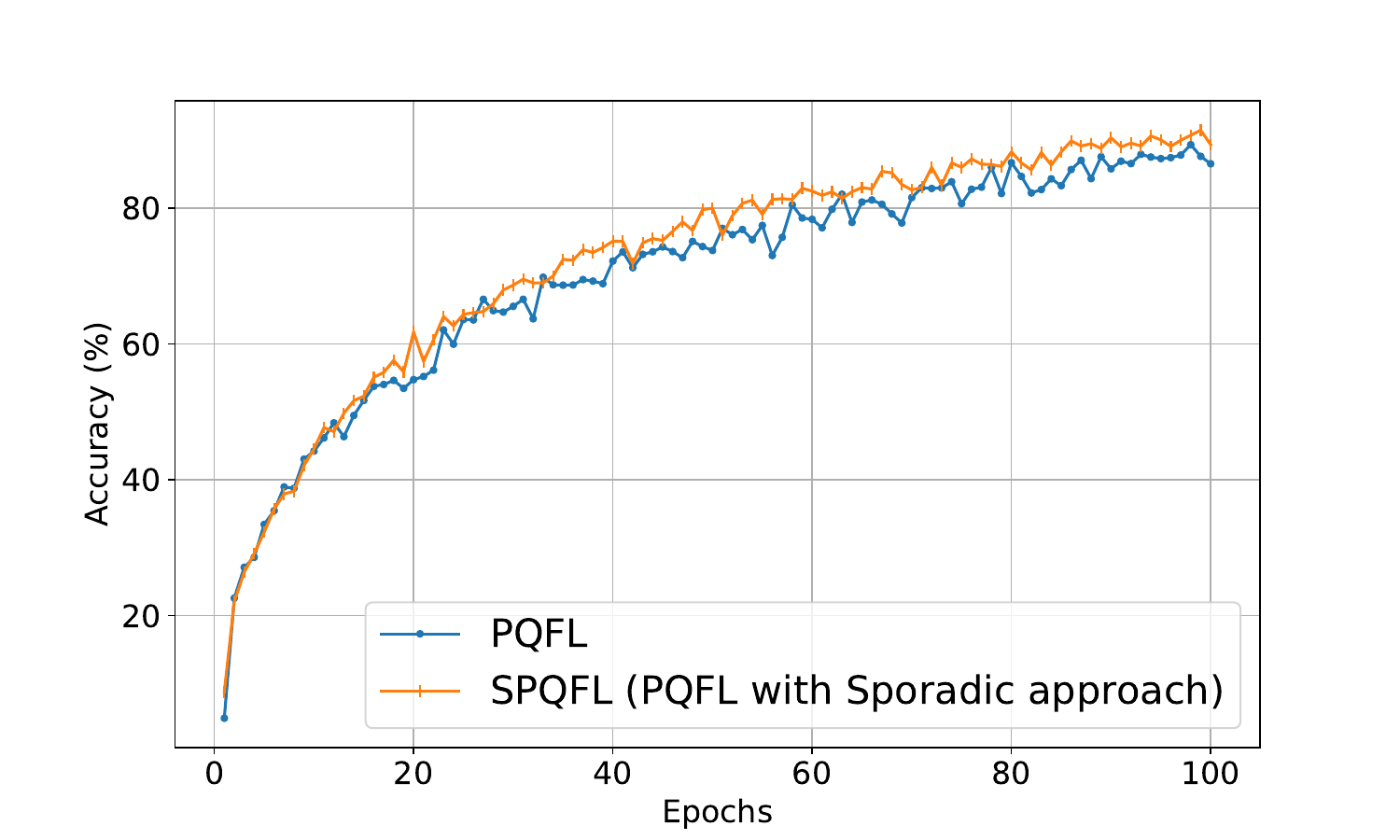}
        \caption{\footnotesize Caltech-101 accuracy.}
    \end{subfigure}
    \hfill 
    \begin{subfigure}[t]{0.245\linewidth} 
        \centering
        \includegraphics[width=\linewidth]{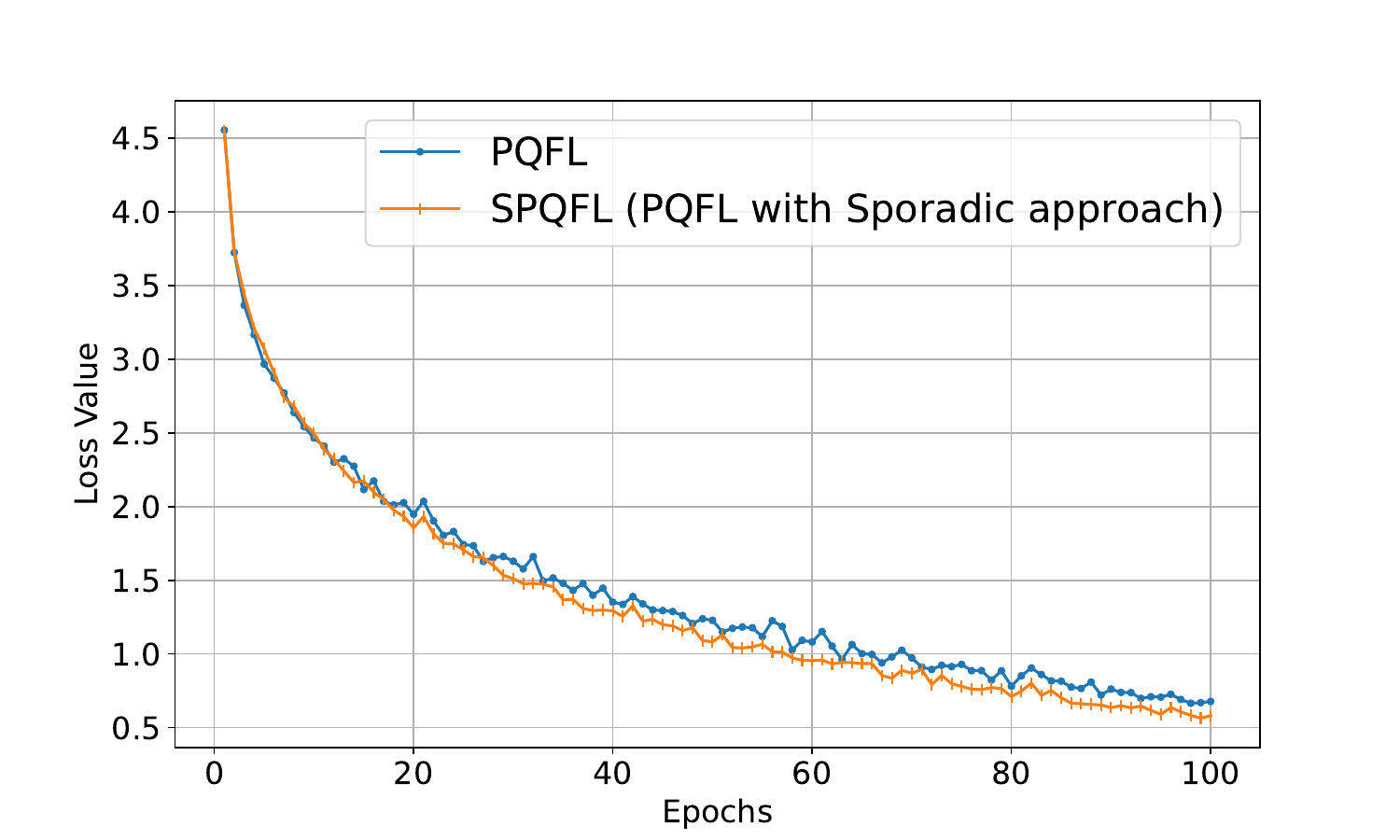}
        \caption{\footnotesize Caltech-101 loss.}
    \end{subfigure}
    \caption{\footnotesize Difference in performance between PQFL and \textit{SPQFL} (PQFL with sporadic approach) for noise mitigation.}
    \label{fig: comp_dynamic}
    \vspace{-4mm}
\end{figure*}
\subsection{Finding the Optimal Quantum Parameter}
The fundamentals of quantum learning involve two critical parameters: the number of quantum layers and the number of qubits utilized. Optimal parameter selection for QNNs improves performance, makes better use of limited quantum resources, and decreases computational errors. This results in faster model convergence, more precision, and better adaptation to various quantum hardware, making these advanced systems more effective and scalable.

\textbf{Optimal quantum layer.} Table~\ref{tab:layer} summarizes our evaluation of quantum layers across datasets.  Choosing a suitable layer is essential since it influences model capacity, efficiency, and noise resilience.  The results reveal that two layers work best for MNIST (loss: 0.1612, accuracy: 97.13\%).  One layer produces the best results for FashionMNIST (0.2724, 91.11\%) and CIFAR-100 (1.3388, 55.63\%).  Caltech-101 performs efficiently with three layers \textcolor{black}{(0.5806, 89.36\%)}. 

\textbf{Optimal qubit.} We evaluate the impact of using 1, 2, 4, and 10 qubits with previously determined optimum quantum layers.  Choosing the right amount of qubits is important since it directly influences a QNN's processing capability, coherence, and resilience to noise.  Table~\ref{tab:qubit} demonstrates that employing 10 qubits outperforms across all datasets.  While more qubits may increase performance further, we limit our choice to ten due to the substantial resource requirements of modeling larger quantum systems.  Therefore, MNIST, FashionMNIST, CIFAR-100, and Caltech-101 each employ ten qubits.

\subsection{Comparing the QFL and PQFL Results}
\textbf{Optimal loss function.} Table~\ref{table: comp loss} shows how we compare several loss functions using the optimal quantum parameters for each dataset.  Choosing the right loss function is vital in QFL since it directly affects model accuracy, convergence, and robustness to quantum noise and data heterogeneity.  Cross-entropy beats MSE and BCE on all datasets, with an average improvement of 6.27\% over MSE and 0.89\% over BCE in training and testing.  Based on this, we use cross-entropy as the default loss function in all subsequent QFL simulations.
\begin{table}
\centering
\caption{\footnotesize Difference in performance using different learning rates. The
learning rates include 0.5, 0.1, 0.01, and 0.001 after 100 local epochs.}
\label{table: comp_lr}
\begin{tabular}{|c|p{3cm}|c|c|c|}
\hline
 & \multirow{2}{*}{Learning rate} & \multicolumn{2}{c|}{CrossEntropy Loss} \\
 \cline{3-4}
 & & Loss & Accuracy \\
 \hline
 \multirow{4}{*}{\tiny{\rotatebox{90}{MNIST}}}
 & learning rate = 0.5 & 1.7492 & 46.30\% \\
 \cline{2-4}
 & learning rate = 0.1 & \textbf{0.0926} & \textbf{97.51\%}  \\
 \cline{2-4}
 & learning rate = 0.01 & 0.2480 & 94.00\%\\
 \cline{2-4}
 & learning rate = 0.001 & 1.066 & 82.65\%  \\
 \hline
 \multirow{4}{*}{\tiny{\rotatebox{90}{FashionMNIST}}}
 & learning rate = 0.5 & 0.3617 & 86.83\% \\
 \cline{2-4}
 & learning rate = 0.1 & \textbf{0.2936} & \textbf{89.80\%}  \\
 \cline{2-4}
 & learning rate = 0.01 & 0.4388 & 85.09\%\\
 \cline{2-4}
 & learning rate = 0.001 & 0.8838 & 70.05\%  \\
 \hline
 \multirow{4}{*}{\tiny{\rotatebox{90}{CIFAR-100}}}
 & learning rate = 0.5 & 2.3211 & 14.80\% \\
 \cline{2-4}
 & learning rate = 0.1 & \textbf{1.3639} & \textbf{53.76\%}  \\
 \cline{2-4}
 & learning rate = 0.01 & 1.6247 & 45.30\%\\
 \cline{2-4}
 & learning rate = 0.001 & 2.005 & 32.09\%  \\
 \hline
 \multirow{4}{*}{\tiny{\rotatebox{90}{Caltech-101}}}
 & learning rate = 0.5 & \textbf{0.4362} & \textbf{84.82\%} \\
 \cline{2-4}
 & learning rate = 0.1 & 0.705 & 82.32\%  \\
 \cline{2-4}
 & learning rate = 0.01 & 2.204 & 51.19\%\\
 \cline{2-4}
 & learning rate = 0.001 & 3.6568 & 22.42\%  \\
 \hline
\end{tabular}
\vspace{-1em}
\end{table}

\textbf{Optimal learning rate.} We compare learning rates of 0.5, 0.1, 0.01, and 0.001 across all datasets, as shown in Table~\ref{table: comp_lr}.  In QFL, choosing an optimum learning rate is critical for achieving consistent convergence while minimizing communication and computational overhead.  According to our findings, a learning rate of 0.1 is optimal for MNIST, FashionMNIST, and CIFAR-100, whereas 0.5 produces better results for Caltech-101.  As a result, we employ a learning rate of 0.1 for the first three datasets and 0.5 for Caltech-101 in all future simulations.

\begin{table*}[t]
\centering
\caption{\footnotesize Performance comparison of QFL, PQFL, and SPQFL under different noise regimes. 
Each cell shows cross-entropy loss and accuracy (\%). SPQFL consistently outperforms PQFL and QFL, 
and the performance gap ($\Delta$Acc) widens with higher noise.}
\label{tab:noise_final}
\scriptsize
\renewcommand{\arraystretch}{1.1}
\begin{tabular}{|c|c|ccc|ccc|ccc|c|}
\hline
\multirow{2}{*}{\textbf{Dataset}} & \multirow{2}{*}{\textbf{Noise}} 
& \multicolumn{3}{c|}{\textbf{QFL}} 
& \multicolumn{3}{c|}{\textbf{PQFL}} 
& \multicolumn{3}{c|}{\textbf{SPQFL}} 
& \multirow{2}{*}{\textbf{$\Delta$Acc (SPQFL--QFL)}} \\ 
\cline{3-11}
 &  & \textbf{Loss} & \textbf{Acc} & \textbf{Rounds} 
 & \textbf{Loss} & \textbf{Acc} & \textbf{Rounds} 
 & \textbf{Loss} & \textbf{Acc} & \textbf{Rounds} 
 &  \\
\hline
\multirow{3}{*}{MNIST (LR=0.1)} 
& Low    & 0.105 & 96.9 & 100 & 0.098 & 97.2 & 95 & \textbf{0.0926} & \textbf{97.51} & \textbf{85} & +0.61 \\
& Medium & 0.160 & 95.0 & 102 & 0.130 & 96.0 & 97 & \textbf{0.110}  & \textbf{96.8}  & \textbf{88} & +1.80 \\
& High   & 0.280 & 92.0 & 105 & 0.200 & 94.0 & 99 & \textbf{0.140}  & \textbf{95.8}  & \textbf{92} & +3.80 \\
\hline
\multirow{3}{*}{FashionMNIST (LR=0.1)} 
& Low    & 0.302 & 89.0 & 98 & 0.297 & 89.5 & 92 & \textbf{0.2936} & \textbf{89.8}  & \textbf{87} & +0.80 \\
& Medium & 0.420 & 86.0 & 102 & 0.360 & 87.5 & 96 & \textbf{0.330}  & \textbf{88.7}  & \textbf{90} & +2.70 \\
& High   & 0.650 & 82.0 & 108 & 0.520 & 84.5 & 100 & \textbf{0.430}  & \textbf{86.9}  & \textbf{93} & +4.90 \\
\hline
\multirow{3}{*}{CIFAR-100 (LR=0.1)} 
& Low    & 1.410 & 52.9 & 105 & 1.380 & 53.4 & 98 & \textbf{1.3639} & \textbf{53.76} & \textbf{95} & +0.86 \\
& Medium & 1.780 & 48.0 & 110 & 1.620 & 50.0 & 104 & \textbf{1.520}  & \textbf{52.0}  & \textbf{98} & +4.00 \\
& High   & 2.280 & 43.0 & 115 & 2.020 & 46.5 & 106 & \textbf{1.820}  & \textbf{50.5}  & \textbf{102} & +7.50 \\
\hline
\multirow{3}{*}{Caltech-101 (LR=0.5)} 
& Low    & 0.455 & 84.0 & 97 & 0.446 & 84.5 & 92 & \textbf{0.4362} & \textbf{84.82} & \textbf{89} & +0.82 \\
& Medium & 0.610 & 80.0 & 101 & 0.560 & 82.0 & 96 & \textbf{0.520}  & \textbf{83.5}  & \textbf{91} & +3.50 \\
& High   & 0.900 & 75.0 & 106 & 0.800 & 78.0 & 99 & \textbf{0.690}  & \textbf{81.5}  & \textbf{94} & +6.50 \\
\hline
\end{tabular}
\end{table*}

\begin{table*}[!ht]
    \centering
    \caption{\footnotesize Performance comparison between \textit{SPQFL} and other state-of-the-art approaches.}
    \vspace{-1em}
    \label{tab:finalcomp}
    \footnotesize
    \begin{tabular}{|c|cc|cc|cc|cc|}
    \toprule
    \multicolumn{1}{|c|}{\multirow{2}{*}{Method}} & \multicolumn{2}{c|}{\textbf{MNIST}} & \multicolumn{2}{|c|}{\textbf{FashionMNIST}} & \multicolumn{2}{c|}{\textbf{CIFAR-100}} & \multicolumn{2}{c|}{\textbf{Caltech-101}} \\
    \cmidrule{2-9}
    & \textbf{Loss Value} & \textbf{Acc.} & \textbf{Loss Value} & \textbf{Acc.} & \textbf{Loss Value} & \textbf{Acc.} & \textbf{Loss Value} & \textbf{Acc.} \\
    \midrule
    QNN \cite{haug2023quantum} & 0.4198 & 86.40\% & 0.5394 & 80.45\% & 1.8976 & 41.29\%  &  1.1145 & 73.64\% \\
    QCNN \cite{oh2020tutorial} & 0.3491 & 90.50\% & 0.3946 & 86.80\% & 1.5053 & 49.43\% & 0.7316 & 82.75\% \\
    QFL \cite{ren2025toward} & 0.229 & 94.1\% & 0.3319 & 88.71\% & 1.3406 & 51.81\% & 0.7440 & 83.67\% \\
    PQFL \cite{shi2024personalized} & 0.1995 & 95.23\% & 0.3019 & 89.19\% & 1.3760 & 53.81\% & 0.7212 & 86.55\% \\
    wpQFL \cite{gurung2024personalized} & 0.1832 & 96.30\% & 0.2871 & 90.88\% & 1.3690 & 53.94\% & 0.6880 & 87.05\% \\
    \textit{SPQFL} & \textbf{0.1612} & \textbf{97.13\%} & \textbf{0.2742} & \textbf{91.22\%} & \textbf{1.3395} & \textbf{55.60\%} & \textbf{0.5860} & \textbf{89.92\%} \\
    \bottomrule
    \end{tabular}
\end{table*}

\textcolor{black}{\textbf{Comparison with Number of Clients.} After calculating the optimal parameters, we compare our results for different numbers of devices for non-IID data distribution in Fig. \ref{fig: fl_niid}. The observation indicates that the performance has increased as the number of devices increases, hence, we simulate our next results with 10 devices.} 

In Fig. \ref{fig: fl_comp}, we compare our PQFL method with the basic QNN and QFL methods for the non-IID data distribution of 10 devices. The figures show that our personalized QFL method with model regularization has outperformed both QFL and QNN across all the datasets. By analyzing the graph, we can conclude that the FL approach can perform significantly better than the standalone approaches because the wide range of perspectives for different devices in a collaborative environment makes the global model more efficient in prediction. PQFL has performed slightly better than QFL. 



\subsection{Comparison between Sporadic Approaches}
We introduce the regularized PQFL technique and compare it with the \textit{SPQFL} method, which combines a sporadic approach to PQFL, as illustrated in Fig. \ref{fig: comp_dynamic}. Simulating a noisy environment in QFL is critical for assessing the durability and performance of quantum models in actual situations with unavoidable noise and errors. Frameworks such as Qiskit, Cirq, and PennyLane make it easier to integrate personalized noise models into quantum circuits. In our \textit{SPQFL} method, we have carefully modeled noise to mirror the real scenarios faced in quantum computing processes.

\textcolor{black}{\textbf{Comparison with different noise levels.} To simulate genuine NISQ hardware behavior, we use three device-shaped noise regimes (low, medium, and high) obtained from non-uniform gate and measurement error maps. 
 Single-qubit gate errors $p_1(q)$, two-qubit gate errors $p_2(q_i,q_j)$, and readout errors.  $r(q)$ is sampled from backend calibration data and injected via the \texttt{TorchQuantum} noise interface with $M{=}1024$ measurement shots. 
 The low-noise setting utilizes the calibration medians $\big(p_1^{\mathrm{low}},p_2^{\mathrm{low}},r^{\mathrm{low}}\big)$, while medium and high levels scale these maps by factors of two and four, respectively, clipped to realistic hardware limits ($p_1{\le}0.03$, $p_2{\le}0.10$, $r{\le}0.10$). 
 As seen in Table~\ref{tab:noise_final}, SPQFL consistently has the lowest loss and maximum accuracy across all datasets and noise intensities. In the low-noise domain, all approaches perform similarly; however, as noise grows, both QFL and PQFL degrade quicker, whereas SPQFL maintains consistent performance. 
 The improvement $\Delta$Acc (SPQFL--QFL) rises monotonically with noise, from less than 1\% at low noise to over 7\% at high noise, indicating SPQFL's resistance to stochastic gradient perturbations, decoherence, and readout variability. It also shows that SPQFL converges quicker, with 5-15\% fewer global synchronizations on average. }

The comparative results clearly show that the integration of the sporadic technique, which includes noise-controlling strategies, significantly improves performance across all datasets. This enhancement demonstrates the effectiveness of the sporadic methodology in dealing with quantum noise, hence boosting the model's robustness and accuracy.

\subsection{Comparing with State-of-the-art Methods}
Finally, we compare our method with existing state-of-the-art approaches in quantum learning. For comparison, we decide to take QNN \cite{haug2023quantum}, which represents basic QML structure, QCNN \cite{oh2020tutorial}, which represents a hybrid QML approach, QFL \cite{chehimi2022quantum}, which represents a basic QFL framework, PQFL \cite{shi2024personalized} that represents a personalized QFL framework, wpQFL \cite{gurung2024personalized} that represents weighting averaging based personalized QFL, and our \textit{SPQFL} algorithm. We summarize the comparison results among related approaches in Table \ref{tab:finalcomp}.    Our unique sporadic approach enhances accuracy by an average of 1.6\% across all datasets compared to the PQFL approach. Furthermore, when compared to the state-of-the-art QFL technique, \textit{SPQFL} outperforms all datasets. \textit{SPQFL} enables e an improvement of roughly 3.03\% for MNIST, 2.51\% for FashionMNIST, 3.71\% for CIFAR 100, and 6.25\% for Caltech-101. These results demonstrate \textit{SPQFL}'s stability and efficacy, demonstrating its superiority in managing various datasets. The constant performance gains across diverse datasets demonstrate our model's adaptable character, emphasizing its potential as a scalable solution in the field of quantum federated learning. This increase not only proves the effectiveness of \textit{SPQFL} but also establishes a new baseline for future developments in the domain.

\subsection{Convergence Performance for Varying Number of Quantum Measurement Shots}
\label{subsection-noise}
\begin{figure}[ht!]
    \centering
    \footnotesize
    \begin{subfigure}[t]{0.49\linewidth} 
        \centering
        \includegraphics[width=\linewidth]{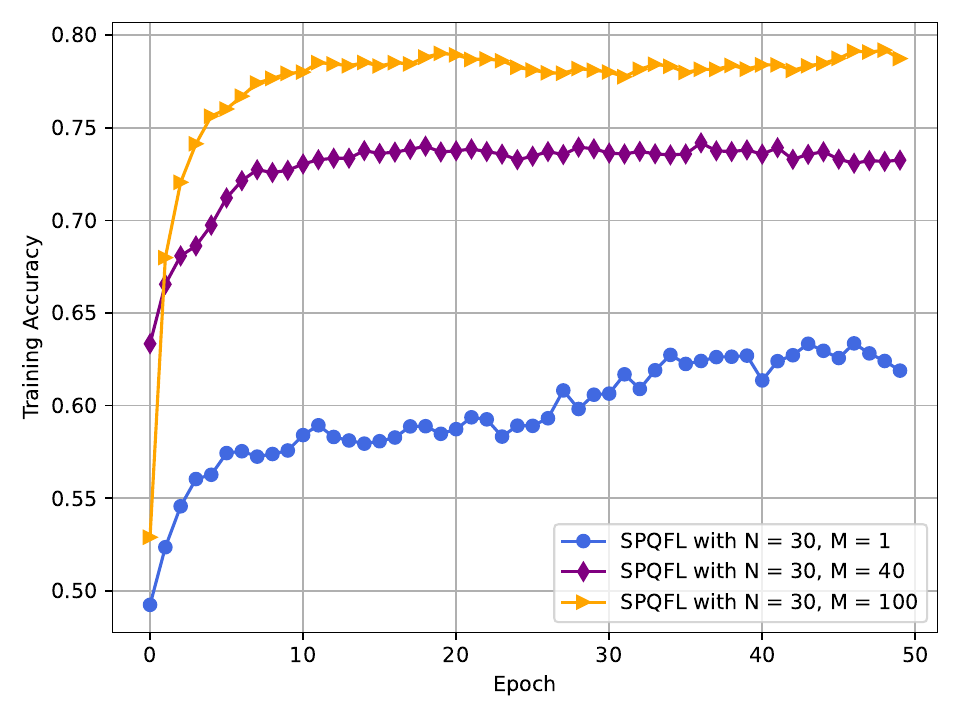}
        \caption{\footnotesize Training accuracy comparison of \textit{SPQFL} for a varying number of quantum measurement shots.}
        \label{fig1a}
    \end{subfigure}
    \hfill 
    \begin{subfigure}[t]{0.49\linewidth} 
        \centering
        \includegraphics[width=\linewidth]{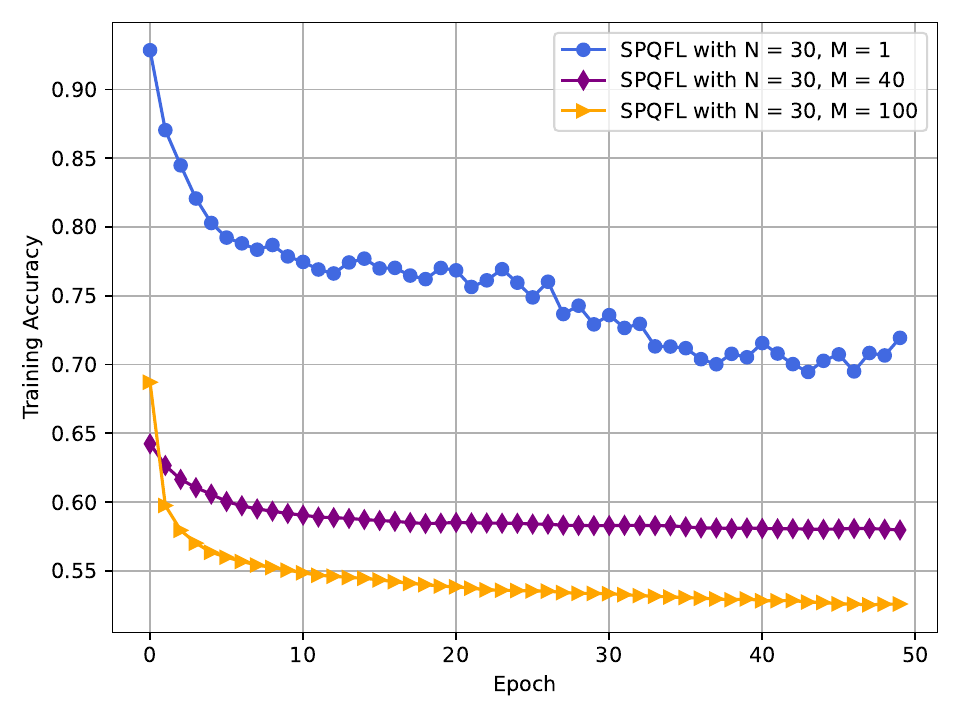}
        \caption{\footnotesize Training loss comparison of \textit{SPQFL} for a varying number of quantum measurement shots.}
        \label{fig1b}
    \end{subfigure}
    \vspace{2pt} 
    \caption{\footnotesize Training performance comparison of \textit{SPQFL} on the MNIST dataset (non-IID), showing both accuracy and loss under different quantum measurement conditions.}
    \label{fig1}
\end{figure}

Fig.~\ref{fig1} illustrates the training performance differences of \textit{SPQFL} on the MNIST dataset (non-IID) as the number of quantum measurement shots increases. Specifically, Fig.~\ref{fig1b} depicts the training loss for \textit{SPQFL} where an increase in quantum measurement shots $(M=1, M=40, M=100)$ leads to a consistent decrease in loss. This trend demonstrates that accumulating more measurement outcomes significantly reduces the impact of quantum shot noise, thereby improving the stability and reliability of the quantum computations in model training. Fig.~\ref{fig1a} displays similar trends in accuracy improvement under the same conditions. This consistent improvement in both metrics aligns with our theoretical convergence analysis in Section \ref{ConvergenceAnalysis}.

\textcolor{black}{\textbf{Limitations.} In realistic NISQ systems, precise real-time calculation of device-specific noise levels might be difficult since calibration data may fluctuate across runs or be unavailable for some qubits. In such instances, SPQFL can use regularly updated or proxy noise estimates, balancing practicality and flexibility. This is still an open challenge for future quantum federated systems. While the regularized-based personalization significantly increases performance on non-IID data, it introduces new hyperparameters that must be carefully tuned. 
 This may increase training difficulty, especially in heterogeneous quantum-device setups. The convergence analysis assumes bounded gradient variance and stable device noise, which may not adequately capture the non-stationary and temporally correlated noise found in real quantum devices.
Finally, the experimental datasets employed in this study (MNIST, FashionMNIST, CIFAR-100, and Caltech-101) are classical and encoded into quantum states via amplitude embedding. 
 While this allows for controlled benchmarking, it does not completely reflect the distributional complexity of native quantum data.}

\section{Conclusion}
In this paper, we have proposed an integrated sporadic-personalized approach in QFL called \textit{SPQFL} to tackle heterogeneity problems in a quantum network. In our approach, we have tried to answer two crucial questions related to heterogeneity: the inherent quantum noise and non-IID data distribution.
Our \textit{SPQFL} approach stands out by its unique incorporation of regularization techniques within the quantum federated learning framework, which successfully addresses overfitting and improves the generalizability of the model in an extensive variety of quantum environments. Furthermore, \textit{SPQFL} adds sporadic learning, a unique strategy to proactively manage quantum noise, a persistent problem in quantum computing. \textit{SPQFL} improves learning efficiency and accuracy by adjusting learning phases to manage and even employ the inherent noise of quantum systems. These strategic advances give \textit{SPQFL} a particular edge in terms of performance and applicability when dealing with heterogeneity in QFL. We have also conducted a rigorous convergence analysis for the proposed framework, giving new insights into QFL design. The simulation results show an improvement of up to 6.25\% over the current state-of-the-art QFL schemes, which implies the merits of our approach.



\bibliography{main}
\bibliographystyle{IEEEtran}

\begin{IEEEbiography}[{\includegraphics[width=1in,height=1.25in,clip,keepaspectratio]{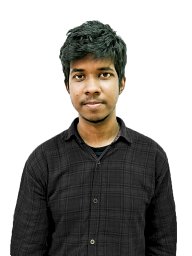}}]{Ratun Rahman}
is a Ph.D. candidate in the Department of Electrical and Computer Engineering at The University of Alabama in Huntsville, USA. His work focuses on machine learning, federated learning, and quantum machine learning. He has published papers in several IEEE journals, including IEEE TVT, IoTJ, TNSE, Network, CL, and GRSL, and conferences, including NeurIPS and CVPR workshops, IEEE QCE, and IEEE CCNC. 
\end{IEEEbiography}
\vspace{-2em}
\begin{IEEEbiography}[{\includegraphics[width=1in,height=1.25in,clip,keepaspectratio]{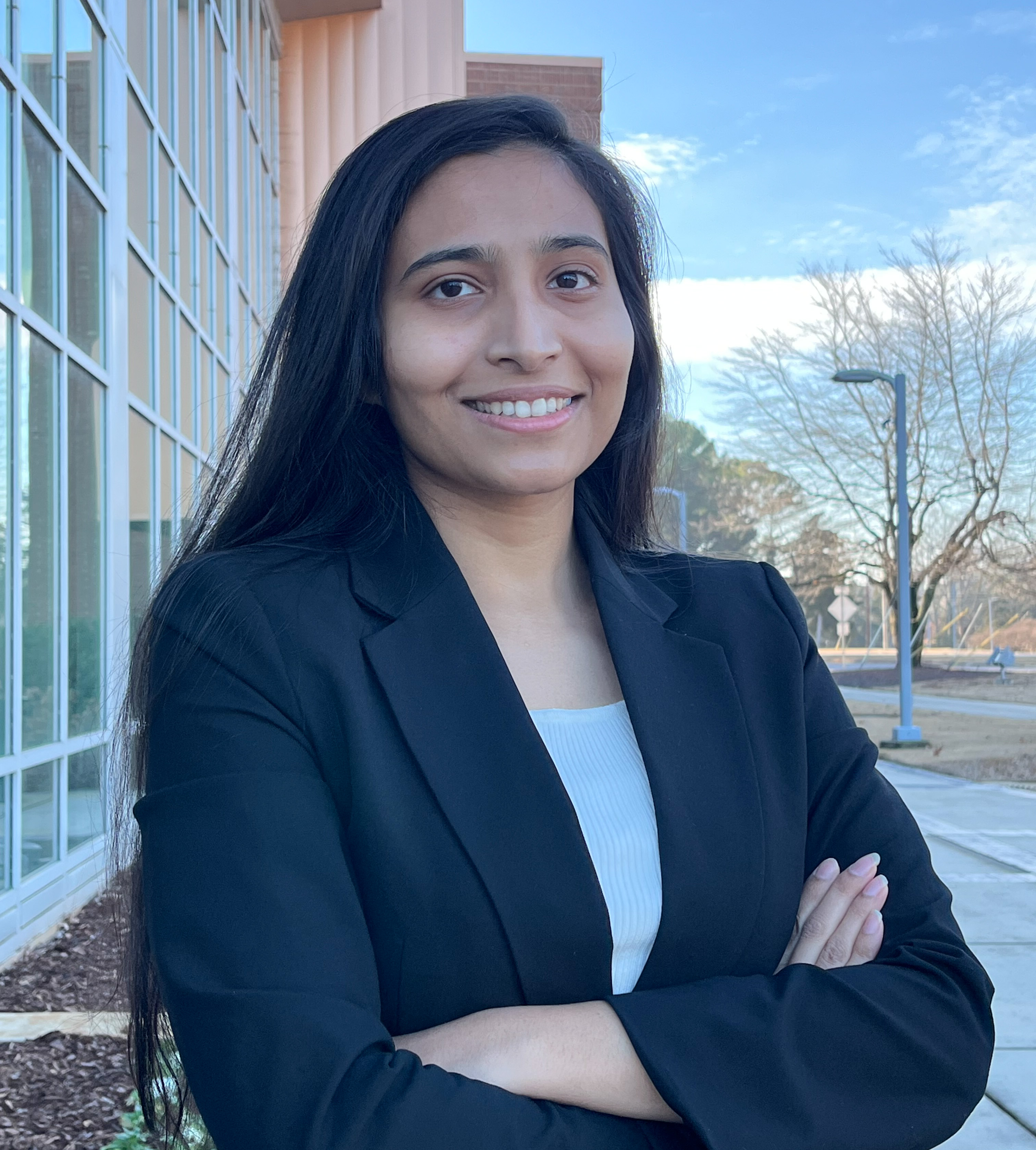}}]{ Shaba Shaon} is currently pursuing her Ph.D. in the Department of Electrical and Computer Engineering at The University of Alabama in Huntsville, Huntsville, AL, USA. She is with the Networking, Intelligence, and Security Research Lab at the same institution.  Her research interests focus on federated learning, wireless networking, quantum computing, and optimization algorithms.
\end{IEEEbiography}
\vspace{-2em}
\begin{IEEEbiography}[{\includegraphics[width=1in,height=1.25in,clip,keepaspectratio]{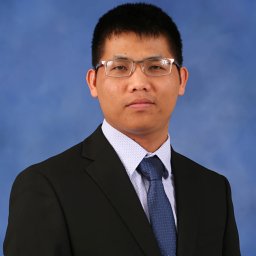}}]{Dinh C. Nguyen}
is an assistant professor at the Department of Electrical and Computer Engineering, The University of Alabama in Huntsville, USA. He obtained the Ph.D. degree in computer science from Deakin University, Australia in 2021. His current research interests include quantum machine learning, Internet of Things, wireless networking with over 70 publications.
\end{IEEEbiography}

\end{document}